\documentclass[letterpaper, 10 pt, conference]{ieeeconf}  

\IEEEoverridecommandlockouts                              

\usepackage{cite}
\usepackage{subfigure}
\usepackage{url}
\usepackage{verbatim,epsf}
\usepackage[font=footnotesize]{subfig}

\usepackage{multicol}
\usepackage{multirow}
\usepackage[font=footnotesize,labelfont=bf]{
	caption}

\usepackage{siunitx}
\usepackage{bm}

\usepackage{tabularx}
\usepackage{balance} 

\usepackage{amssymb}
\usepackage{amsmath}
\usepackage{graphicx}
\usepackage{epstopdf}

\usepackage{epsfig, color}

\usepackage{comment} 
\usepackage{amsfonts} 
\usepackage{amsmath} 
\usepackage{rotating}

\usepackage{amssymb}
\usepackage{amsmath}
\usepackage{amsfonts}
\usepackage{wasysym}

\def\QED{\mbox{\rule[0pt]{1.5ex}{1.5ex}}}

\usepackage{algorithm}
\usepackage{algorithmic}

\usepackage{color}

\title{\LARGE \bf
DeepPoint: A Deep Learning Model for 3D Reconstruction in Point Clouds via mmWave Radar
}

\author{Yue Sun$^{1}$, Honggang Zhang$^{2}$, Zhuoming Huang$^{3}$, and Benyuan Liu$^{4}$
\thanks{$^{1}$Yue Sun is in the Department of Computer Science, UMass Boston, USA.
        {\tt\small Yue.Sun001@umb.edu}}%
\thanks{$^{2}$Honggang Zhang is in the Department of Engineering, UMass Boston, USA.
	{\tt\small Honggang.Zhang@umb.edu}}%
\thanks{$^{3}$Zhuoming Huang is in the Department of Engineering, UMass Boston, USA.
	{\tt\small Zhuoming.Huang001@umb.edu}}%
\thanks{$^{4}$Benyuan Liu is in the Department of Computer Science, UMass Lowell, USA.
	{\tt\small bliu@cs.uml.edu}}%
}

\begin{document}

\maketitle
\thispagestyle{empty}
\pagestyle{empty}

\begin{abstract}

Recent research has shown that mmWave radar sensing is effective for object detection
in low visibility environments,
which makes it an ideal technique in autonomous navigation systems such as autonomous vehicles.
However, due to the characteristics of radar signals such as 
sparsity, low resolution, specularity, and high noise,
it is still quite challenging to reconstruct 3D object shapes via mmWave radar sensing.
Built on our recent proposed 3DRIMR (\textbf{3D} \textbf{R}econstruction and \textbf{I}maging via 
\textbf{m}mWave \textbf{R}adar), we introduce in this paper DeepPoint, a deep learning model 
that generates 3D objects in point cloud format that 
significantly outperforms the original 3DRIMR design.
The model adopts a conditional Generative Adversarial Network (GAN) based deep neural network architecture.
It takes as input the 2D depth images of an object generated by 3DRIMR's Stage 1, and  
outputs smooth and dense 3D point clouds of the object.
The model consists of a novel generator network that utilizes
a sequence of DeepPoint blocks or layers to extract
essential features of the union of multiple rough and sparse 
input point clouds of an object when observed from
various viewpoints,
given that those input point clouds may contain many incorrect points due to the imperfect generation
process of 3DRIMR's Stage 1.  
The design of DeepPoint adopts a deep structure 
to capture the global features of input point clouds,
and it relies on an optimally chosen number 
of DeepPoint blocks and skip connections to achieve 
performance improvement over the original 3DRIMR design. 
Our experiments have demonstrated that this model significantly outperforms the original 
3DRIMR and other standard techniques in reconstructing 3D objects.

\end{abstract}

\section{INTRODUCTION}

The advantage of Millimeter Wave (mmWave) radar in object sensing in low visibility environment has 
been actively studied recently in applying it in autonomous 
vehicles \cite{HawkEye} and search/rescue in high risk areas \cite{mobisys20smoke}.
However further application of mmWave radar in object imaging and reconstruction is quite difficult due to 
the characteristics of mmWave radar signals such as low resolution, sparsity, and high noise due to multi-path and 
specularity. Recent work \cite{HawkEye, superrf,mobisys20smoke} attempt to design deep learning systems 
to generate 2D depth images based on mmWave radar signals. 3DRIMR \cite{sun20213drimr} further introduces 
a design that generates 3D object shapes based on mmWave radar, but the end results are still not satisfactory. 

In this paper, we introduce DeepPoint, a deep learning model 
that generates 3D objects in dense and smooth 
point clouds based on the union of multiple rough and sparse 
input point clouds, which are directly converted from  
the 2D depth images generated by the Stage 1 of 3DRIMR which takes raw radar data as input. 
The training of DeepPoint follows conditional GAN architecture, and it significantly outperforms the original 3DRIMR design.

The 3DRIMR \cite{sun20213drimr} architecture consists of two stages, and each has a generator network. 
Stage 1's generator network $\bf{G_{r2i}}$ takes 3D radar intensity data as input and generates 2D depth images;
Stage 2's generator network $\bf{G_{p2p}}$ takes as input a set of multiple 2D depth images 
and outputs the 3D shape of the object in the format of point cloud.  
Each stage's generator network is jointly trained with a separate discriminator network, using conditional GAN architecture. 
The design rationale for 3DRIMR architecture is to combine the advantages of Convolutional Neural Network (CNN)'s 
convolutional operation and the efficiency of point cloud representation of 3D objects, as the former
can capture detailed local neighborhood structure of a 3D object, 
and the latter is more efficient and of higher resolution than 3D
shape representation via voxelization. 

Even though 3DRIMR has shown promising results, its Stage 2's generator network design is still not quite 
satisfactory. Specifically, the edges of generated 3D objects are still blurry and the points of an object's point 
cloud tend to evenly distributed in space which do not give a clear sharp shape structure. 
DeepPoint introduced in this paper replaces the Stage 2 of the original 3DRIMR, and it significantly outperforms 3DRIMR both quantitatively and visually. 

Our major contributions are as follows: 
\begin{enumerate}
	\item 
	DeepPoint, a novel generator network that can generate smooth and dense point cloud representation of 
	a 3D object based on the union of multiple rough and sparse point clouds directly converted from 
	the 2D depth images derived from raw mmWave radar sensor data. 
	The generator network utilizes
	a sequence of DeepPoint blocks or layers to extract
	essential features of those input point clouds of an object when observed from
	various viewpoints,
	even though those input point clouds may contain many incorrect points due to the imperfect generation
	process of 3DRIMR's Stage 1.  
	
	\item 
	Novel designs such as a conditional GAN architecture design for the training of DeepPoint, 
	an optimally chosen number of layers and skip connection. Those designs have resulted in the
	performance improvement over the original 3DRIMR.

	\item 
	An improved 3DRIMR system implementation that can conduct fast 3D object reconstruction
	by using a commodity 
	mmWave radar sensor, instead of a slow full-scale SAR scan.
	The whole system takes advantage of convolutional operation and point cloud based neural network for efficient 3D shape generation with detailed geometry.  
\end{enumerate}

In the rest of the paper, we briefly discuss related work and preliminaries in Sections 
\ref{sec_related} and \ref{sec_background}.
Then we discuss the design of DeepPoint model in Section \ref{sec_design}. 
Experiment results are given in Section \ref{sec_imp}. 
Finally the paper concludes in Section \ref{sec_conclusion}.

\section{RELATED WORK}\label{sec_related}

Frequency Modulated Continuous Wave (FMCW) Millimeter Wave (mmWave) radar sensing 
has been an active research area in recent years, especially in applications such as person/gesture identification \cite{vandersmissen2018indoor,yang2020mu}, car detection/imaging \cite{HawkEye}, and environment sensing \cite{mobisys20smoke, superrf}. Usually Synthetic Aperture Radar (SAR) is used in data collection 
for high resolution, e.g., \cite{mamandipoor201460,national2018airport,ghasr2016wideband,sheen2007near}.

This paper is built on our recent work \cite{sun20213drimr} 
on applying mmWave radar for 3D object reconstruction, in which we proposed 3DRIMR system. 
The deep neural network model proposed in this paper completely replaces the model in the Stage 2 of 3DRIMR, 
and this new model significantly outperforms the original 3DRIMR. 
There have been a few recent work on mmWave radar based imaging, mapping, and 3D object 
reconstruction \cite{HawkEye,superrf,mobisys20smoke,yuan2018pcn,qi2016pointnet}. Our work is inspired by 
their promising research results, and due to the low cost and small form factor of commodity mmWave 
radar sensors, we plan to develop a simple and fast 3D reconstruction system to 
be attached in our UAV SLAM system \cite{sun2020lidaus} for search and rescue in dangerous environment.

Besides radar signals, vision community has also been working on learning-based 
3D object shape reconstruction \cite{yang20173d,dai2017shape,sharma2016vconv,smith2017improved}, 
most of which use voxels to represent 3D objects.
Our proposed neural network model  
uses point cloud as a format for 3D objects to capture detailed geometric information 
with efficient memory and computation performance. 

PointNet structure is utilized in PCN\cite{yuan2018pcn}, which uses point cloud to reconstruct 3D object shapes, and this structure inspires us to design our model. The novelty of our model is
that it has a deeper structure than PCN and skip connections are used for better capture of 
objects' edges and shapes. In addition, our work adopts a conditional GAN architecture to jointly train a generator and a discriminator for better performance. 

\section{Preliminaries}\label{sec_background}

\subsection{FMCW Millimeter Wave Radar Sensing and Imaging}

Similar to \cite{sun20213drimr}, we use Frequency Modulated Continuous Wave (FMCW) mmWave radar 
sensor \cite{timmwave} signals to reconstruct 3D object shapes. 
Three Fast Fourier Transforms (FFTs) are conducted on received waveforms
to generate 3D heatmaps or intensity maps of the space that
represent the energy or radar intensity per voxel, written as $x(\phi, \theta, \rho)$.
Note that  $\phi$, $\theta$, and $\rho$ represent azimuth angle, elevation angle, and range respectively. 
Same as in \cite{sun20213drimr}, we use IWR6843ISK \cite{iwr6843} operating at $60$ GHz frequency,
and for high resolution radar signals, we adopt the Synthetic Aperture Radar (SAR) operation.
Unlike data from LiDAR and camera sensor, mmWave radar sensors can only give us 
sparse, low resolution, and highly noisy data. Partically, incorrect ghost points in radar signals 
can be generated due to multi-path effect. Reference \cite{HawkEye,superrf,mobisys20smoke} give more detailed discussion 
on FMCW mmWave radar sensing. 

\subsection{Representation of 3D Objects}

In this work, we adopt point cloud format to represent 3D objects. 
Even though point cloud format is a standard representation of 3D objects and it is used in 
learning-based 3D reconstruction, e.g., \cite{fan2017point,qi2016pointnet,qi2017pointnet++}, 
but CNN convolutional operation cannot be directly applied to a point cloud set as it is 
essentially an unordered point set.  Furthermore, the point cloud of an object that is directly generated by 
raw radar signals is not a good choice to reconstruct the object due to the radar signal's 
low resolution, being sparse,  and with incorrect ghost points due to multi-path effect. 
Besides point clouds,  voxel representations can also be used in 3D reconstruction \cite{kar2017learning,paschalidou2018raynet,ji2017surfacenet,wu2016learning}, and the advantage of such 
representation is that 3D CNN convolutional operations can be applied to it. 
In addition, mesh representations of 3D objects are also used in existing work \cite{kong2017using,wang2018pixel2mesh}.
However these two representation formats are limited by memory and computation cost.

\subsection{Review of 3DRIMR Architecture}

This paper introduces DeepPoint as the generator and discriminator networks of the Stage 2 of 3DRIMR to generate
smooth and dense point clouds.
For completeness, we now briefly review 3DRIMR architecture. 

3DRIMR consists of two back-to-back generator networks 
$\mathbf{G_{r2i}}$ and $\mathbf{G_{p2p}}$.
In Stage 1, $\mathbf{G_{r2i}}$ receives a 3D radar energy intensity map
of an object and outputs a 2D depth image of the object.
We let a mmWave radar sensor scans an object from multiple viewpoints
to get multiple 3D energy maps. Then  $\mathbf{G_{r2i}}$ generates 
multiple 2D depth images of the object. 
The Stage 2 of 3DRIMR first pre-processes these images to get multiple coarse point clouds of the object,
which are used as input to $\mathbf{G_{p2p}}$ to generate a single point cloud of 
the object. 
A conditional GAN architecture is designed for 3DRIMR's training. That is,
two discriminator networks $\mathbf{D_{r2i}}$ and $\mathbf{D_{p2p}}$ that are jointly trained together with their corresponding generator networks.

Let $m_r$ denote a 3D radar intensity map of an object captured 
from a viewpoint, and 
let $g_{2d}$ be a ground truth 2D depth image of the same object captured 
from the same viewpoint. 
$\mathbf{G_{r2i}}$ generates $\hat{g}_{2d}$ that predicts or estimates $g_{2d}$ given $m_{r}$. 
If there are $k$ different viewpoints $v_1, ..., v_k$, 
generator $\mathbf{G_{r2i}}$ predicts their corresponding 2D depth images $\{\hat{g}_{2d,i} | i = 1, ..., k\}$.
Each $\hat{g}_{2d,i}$ can be directly converted to a coarse and sparse 3D point cloud. 
Then we can have $k$ coarse point clouds $\{P_{r,i} | i = 1, ..., k\}$
of the object. 
The Stage 2 of 3DRIMR unions the $k$ coarse point clouds 
to form an initial estimated coarse point cloud of the object, denoted as $P_r$, which is 
a set of 3D points $\{p_j | j = 1, ..., n\}$.
Generator $\mathbf{G_{p2p}}$ takes 
$P_r$ as input, 
and predicts a dense, smooth, and accurate point cloud $\hat{P}_r$. 
Note that since the prediction of $\mathbf{G_{r2i}}$ may not be completely correct, a coarse 
$P_r$ may likely contain many missing or even incorrect points. 

The Stage 1's design of 3DRIMR can be found in \cite{sun20213drimr}. Next we discuss our proposed 
DeepPoint as the network model in the Stage 2 of 3DRIMR.

\begin{figure*}[htb!]
	\centerline{
		\begin{minipage}{3.5in}
			\begin{center}
				\setlength{\epsfxsize}{3.4in}
				\epsffile{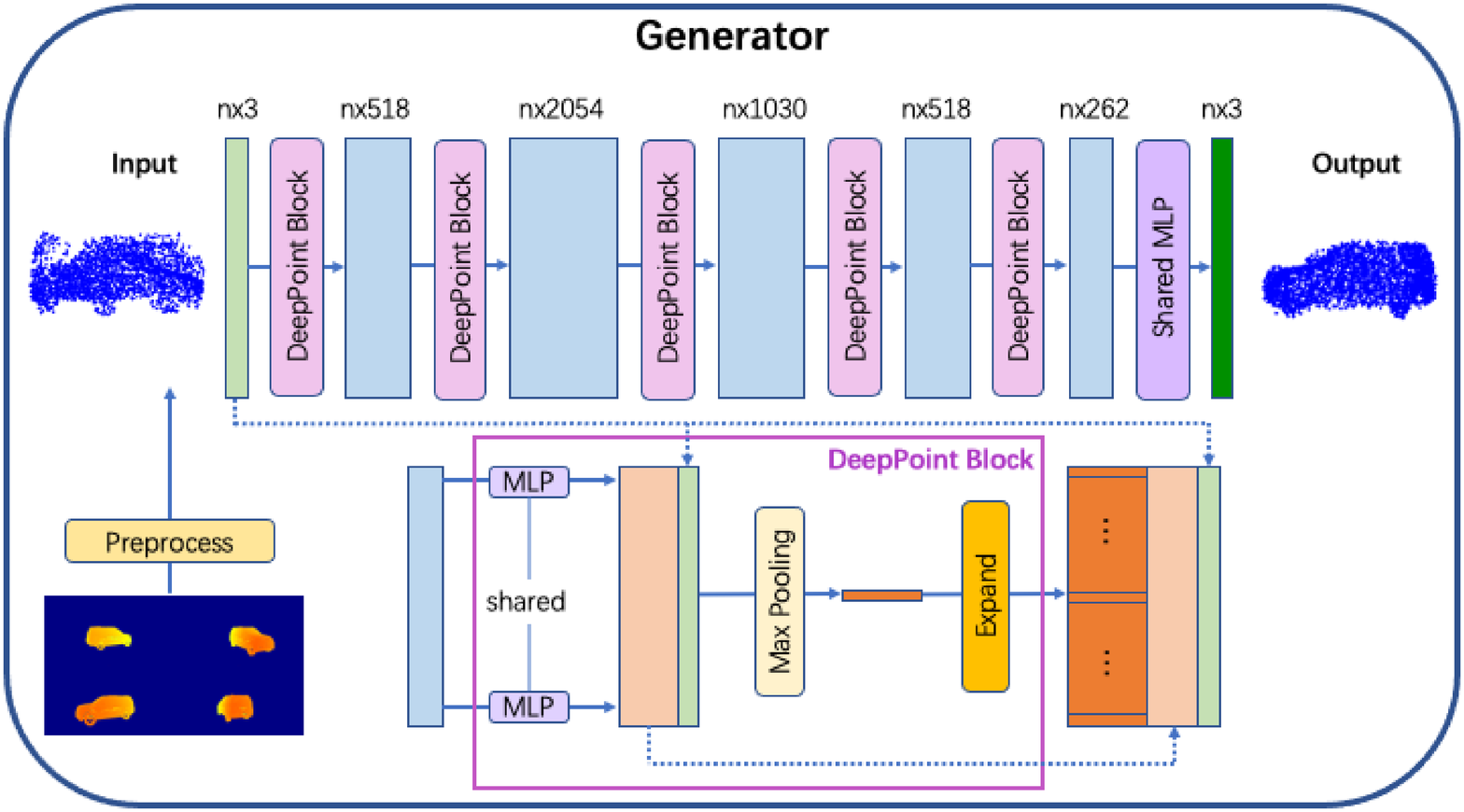}\\
				{}
			\end{center}
		\end{minipage}
		\begin{minipage}{3.5in}
			\begin{center}
				\setlength{\epsfxsize}{3.4in}
				\epsffile{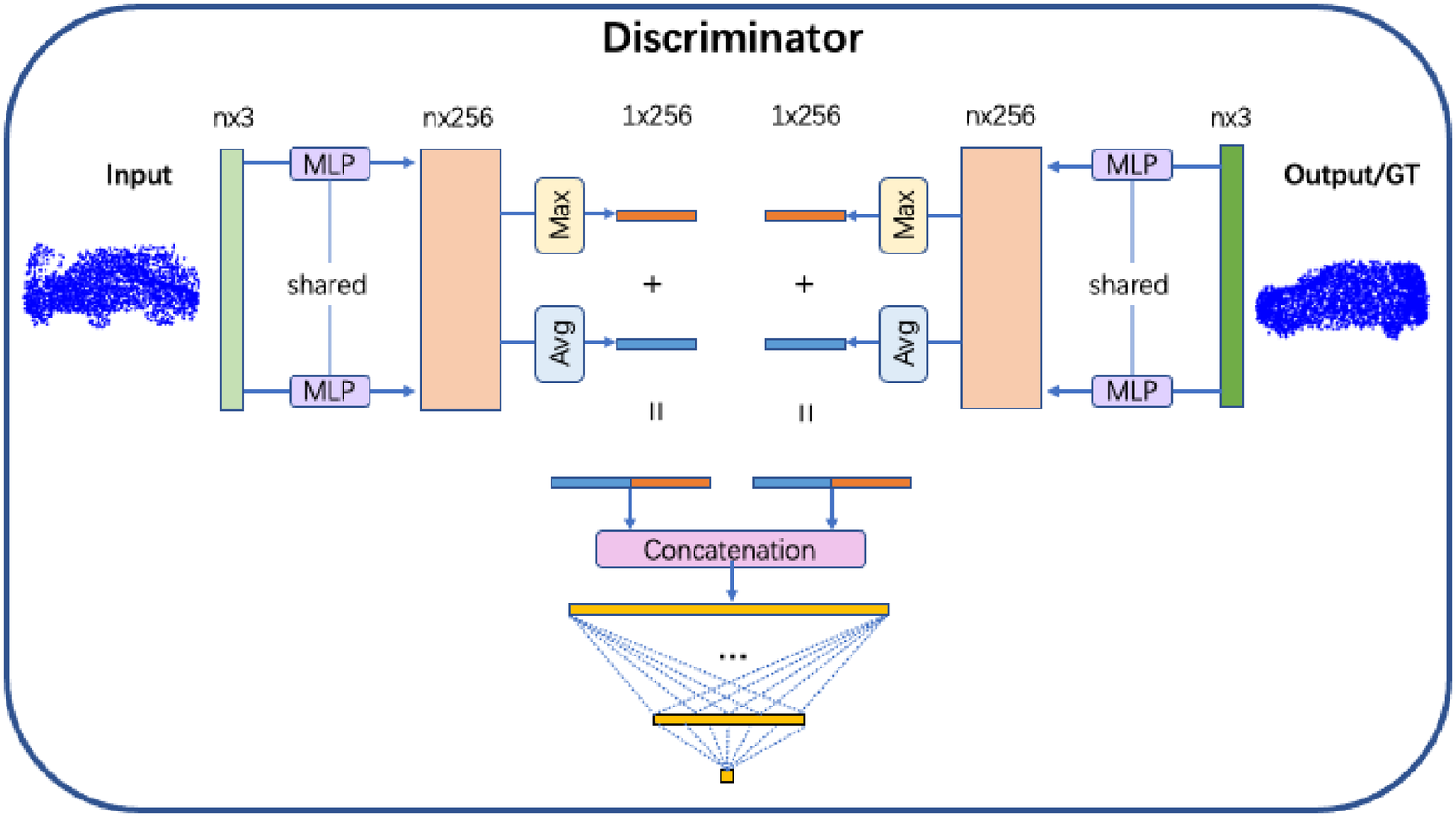}\\
				{}
			\end{center}
		\end{minipage}
	}
	\caption{DeepPoint architecture proposed in this paper. 
		The system first pre-processes multiple 2D depth images of an object   
		to get multiple coarse and sparse point clouds, which may contain many incorrect points. 
		Those depth images are generated for the same object but viewed from four different viewpoints.
		Combining those coarse point clouds we can derive a single coarse point cloud, 
		which is used as the input of the generator $\mathbf{G_{p2p}}$ of DeepPoint.
		The generator of DeepPoint outputs a dense and smooth point cloud representation of the object. 
		A conditional GAN architecture is used to train both generator $\mathbf{G_{p2p}}$ 
		and discriminator $\mathbf{D_{p2p}}$ jointly.}
	\label{fig_stage_2}
\end{figure*}

\section{DeepPoint Design}\label{sec_design}

\subsection{Overview}

DeepPoint is a generative model to generate a smooth and dense 3D point cloud of an object 
from the union of multiple coarse point clouds. which are directly 
converted from 2D depth images of the object observed from different viewpoints.
Since those 2D depth images are generated from raw radar energy maps, the union of
their converted point clouds may contain many incorrect points due to the imperfect 
image generation process. DeepPoint's generator network is able to correct those incorrect
points and generate a smooth and dense point cloud. This generator 
can be used as the network in the Stage 2 of 3DRIMR, and it also can be used as an independent 
network model that works on any rough and sparse input point clouds that contains incorrect points. 

The training of DeepPoint is a conditional GAN architecture,  
as shown in Fig. \ref{fig_stage_2}.
DeepPoint's generator network $\mathbf{G_{p2p}}$ aims at generating a point cloud of an object 
with continuous and smooth contour $\hat{P}_r$, 
from $P_r$, a union of $k$ separate coarse point clouds observed from $k$ viewpoints of the object $\{P_{r,i} | i=1,...,k\}$.

\subsection{Generator $\mathbf{G_{p2p}}$}

Generator $\mathbf{G_{p2p}}$'s input $P_r$ and output $\hat{P}_r = \mathbf{G_{p2p}}(P_r)$ are 3D point clouds represented as $n \times 3$ matrices, 
with each row being the 3D Cartesian coordinate $(x, y, z)$ of a point. 
However, different from the Stage 2's generator in \cite{sun20213drimr}, 
the input and output point clouds in our work 
should have the same number of points due to the use of 
Earth Mover's Distance (EMD) in the loss function design for the training of the generator network.

Generator $\mathbf{G_{p2p}}$ consists of a series of blocks that 
expand an input point cloud $P_r$ into a high dimensional feature 
and then shrink it into 3-dimensional outputs $\hat{P}_r$. These blocks are referred to as 
DeepPoint Block, 
and their designs are inspired by PointNet structure. 
During the training, a discriminator $\mathbf{D_{p2p}}$ takes $(P_r, P_{true})$ or $(P_r, \hat{P}_r)$ 
pairs and outputs a score to indicate the difference between them. 

As shown left part of Fig. \ref{fig_stage_2}, the generator
takes the input $P_r$, and converts it into a high-dimensional point feature matrix after passing 
first 2 DeepPoint blocks, 
and then decreases the dimension of feature matrices through the following 3 DeepPoint blocks, 
and finally outputs a new set of 3-dimensional matrix $\hat{P}_r$,
which represents the Cartesian coordinate $(x, y, z)$ of a point of the predicted point cloud.

The detailed structure of each {DeepPoint} block is shown in the bottom dotted box of Generator in Fig. \ref{fig_stage_2}.
It first passes the raw input into a shared MLP, 
and then concatenates the raw input, i.e., 
the Cartesian coordinate $(x, y, z)$ of the input point cloud,
to form a point feature matrix $F_{P_r}$ with each row representing the local feature of a corresponding point.
Then, it applies a point-wise maxpooling on $F_{P_r}$ and extracts a global feature vector $g_{P_r}$.
To produce a complete point cloud for an object, 
we need both local and global features, 
therefore, we concatenates the global feature $g_{P_r}$ with each of the point features
$f_i$ and form another matrix $F'_{P_r}$.

\bigskip
\noindent \textbf{Remarks.}
Note that there are two major differences between the proposed generator network and 
the generator in Stage 2 of 3DMIMR \cite{sun20213drimr}.
First, the proposed generator is ``deeper" than the generator of 3DMIMR 
as ours has more layers of {DeepPoint} blocks.
Second, 3DMIMR uses fully connected layers and apply reshape operation to derive output point cloud from a high-dimensional global feature vector, 
which can only get an rough overall shape without many fine, local, and detailed characteristics.
However, our new generator network design
generates an output point cloud from both local and global features 
and hence can generate fine, local and detailed characteristics of an object.

\subsection{Discriminator $\mathbf{D_{p2p}}$}


We design an improved discriminator with two-stream inputs, as shown in Fig. \ref{fig_stage_2}.
The inputs pass through a simple shared MLP and get expanded into higher dimensional matrices.
A point-wise max pooling and an average pooling are used to extract two global feature vectors.
Then the two global feature vectors are concatenated to form final global features, 
which are further concatenated and fed into two fully connected layers to derive a score. 
The score is used to indicate whether the input is real or fake, i.e., generated point cloud.

\subsection{Loss Function}

When training the generator network, we concurrently train $\mathbf{D_{p2p}}$ to minimize $\mathcal{L}_{\mathbf{D_{p2p}}}$, 
and train $\mathbf{G_{p2p}}$ to minimize $\mathcal{L}_{\mathbf{G_{p2p}}}$. 
$\mathcal{L}_{\mathbf{D_{p2p}}}$ is calculated as the mean MSE (Mean Square Error) of $\mathbf{D_{p2p}}$'s prediction error.
The loss function of generator $\mathcal{L}_{\mathbf{G_{p2p}}}$ is a weighted sum, 
consisting of $\mathcal{L}_{GAN} (\mathbf{G_{p2p}})$, 
Chamfer loss $\mathcal{L}_{cf}$  
between predicted point clouds and the ground truth, 
and EMD \cite{yuan2018pcn} loss $\mathcal{L}_{emd} (\mathbf{G_{p2p}})$. 

Note that Chamfer distance \cite{yuan2018pcn} calculates the average closest distance between input and output points.
The symmetric version of it is defined as:
\begin{equation}
\small{
	d_{cf}(S_1, S_2) = 
	\frac{1}{|S_1|} \sum_{x \in S_1} \mathop{min}\limits_{y \in S_2} \lVert x-y \rVert_{2} 
	+ \frac{1}{|S_2|} \sum_{y \in S_2} \mathop{min}\limits_{x \in S_1} \lVert y-x \rVert_{2}
}
\end{equation}
Then, Chamfer loss isdefined as:
\begin{equation}
\mathcal{L}_{cf} (\mathbf{G_{p2p}}) = d_{cf}(\hat{P}_r, P_{true})
\end{equation}
In addition, Earth Mover's Distance (EMD) \cite{yuan2018pcn} can find a bijection $\phi: S_1 \to S_2$, 
which can minimize the average distance between pairs of corresponding points. 
The equation of EMD is:
\begin{equation}
\small{
d_{emd}(S_1, S_2) =  
\mathop{min}\limits_{\phi: S_1 \to S_2} \frac{1}{|S_1|}
\sum_{x \in S_1} || x-\phi(x) ||_{2} 
}
\end{equation}
In our case, EMD loss is calculated as:
\begin{equation}
\mathcal{L}_{emd} (\mathbf{G_{p2p}}) = d_{emd}(\hat{P}_r, P_{true})
\end{equation}

$\mathcal{L}_{\mathbf{G_{p2p}}}$ is given by Eqn. (\ref{eqn_L_G_p2p}). 
Note that $\lambda_{d_{cf}}$ and $\lambda_{d_{emd}}$ are hand-tuned to 100 and 1 respectively
in our experiments. 
\begin{eqnarray}\label{eqn_L_G_p2p}
\mathcal{L}_{\mathbf{G_{p2p}}} &=& \mathcal{L}_{GAN} (\mathbf{G_{p2p}})
+ \lambda_{d_{cf}}\mathcal{L}_{cf}(\mathbf{G_{p2p}})  \nonumber \\
&&+ \lambda_{d_{emd}} \mathcal{L}_{emd} (\mathbf{G_{p2p}})
\end{eqnarray}

\section{IMPLEMENTATION AND EXPERIMENTS}\label{sec_imp}

We implement DeepPoint and use it as Stage 2's generator network of 
3DRIMR \cite{sun20213drimr} system. 
The system first generates 2D depth images from 
3D radar intensity maps from multiple views of an object,
and then passes these output depth images to the generator network of DeepPoint
to produce a 3D point cloud of the object.
We now present our experiment results in this section. 

\subsection{Datasets}

\begin{figure}[htb!]
	\centerline{
		\begin{minipage}{2.2in}
			\begin{center}
				\setlength{\epsfxsize}{2.2in}
				\epsffile{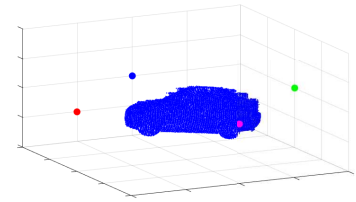}\\
				{}
			\end{center}
		\end{minipage}
	}
	\caption{An example scene of a car.}
	\label{fig_car_ex}
\end{figure}

We conduct experiments on cars with average size of $445cm \times 175cm \times 158cm$.  
The input data to the proposed the generator network is 
the output depth images produced by 3DRIMR's Stage 1.
We follow a procedure that is similar to 3DRIMR \cite{sun20213drimr}, 
to generate ground truth point clouds.
Fig. \ref{fig_car_ex} shows an example scene.

\subsection{Model Training and Testing}

\begin{figure*}[htb!]
	\begin{minipage}{7.0in}
		\begin{center}
			\setlength{\epsfxsize}{1.6in}
			\epsffile{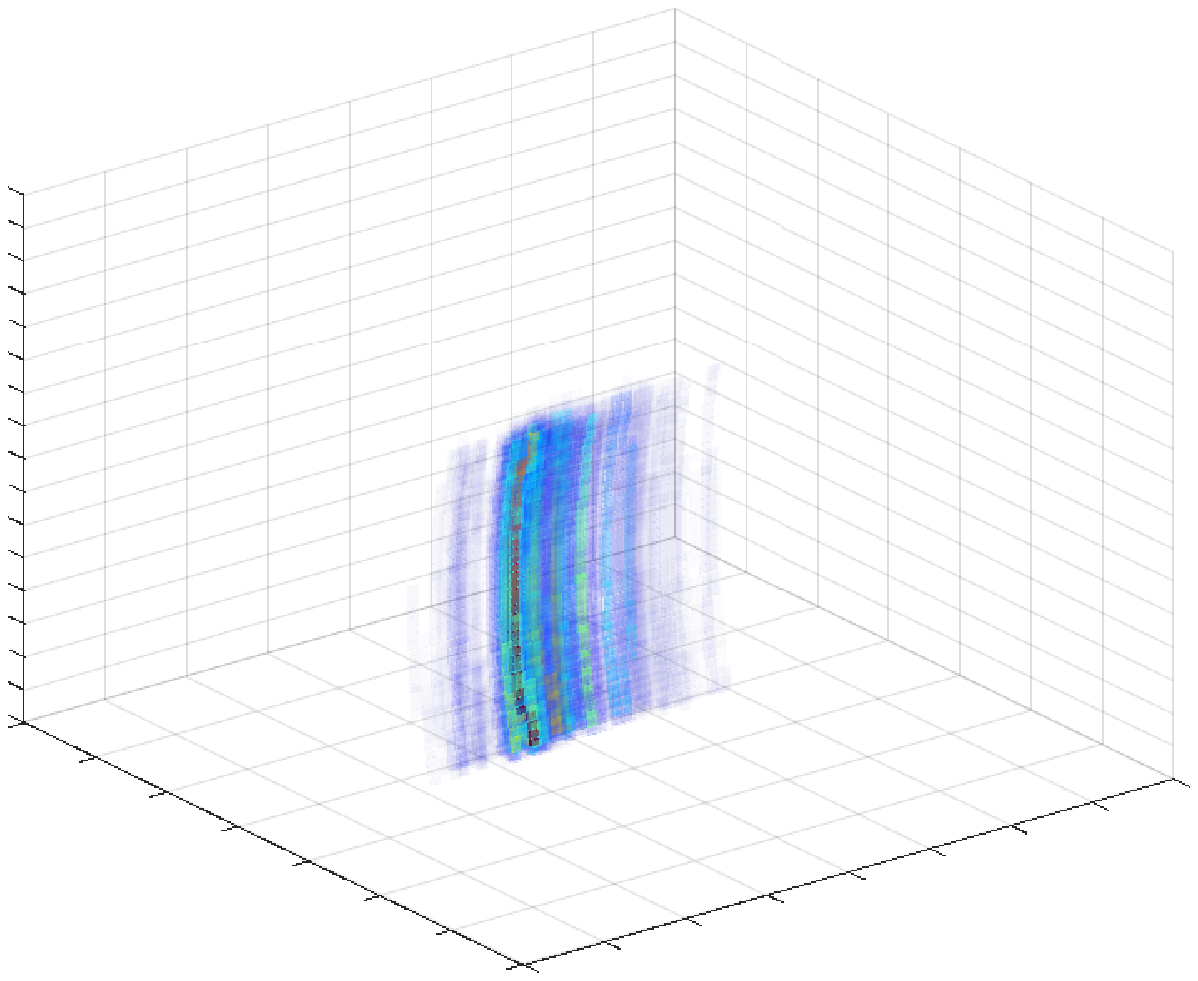}
			\setlength{\epsfxsize}{1.6in}
			\epsffile{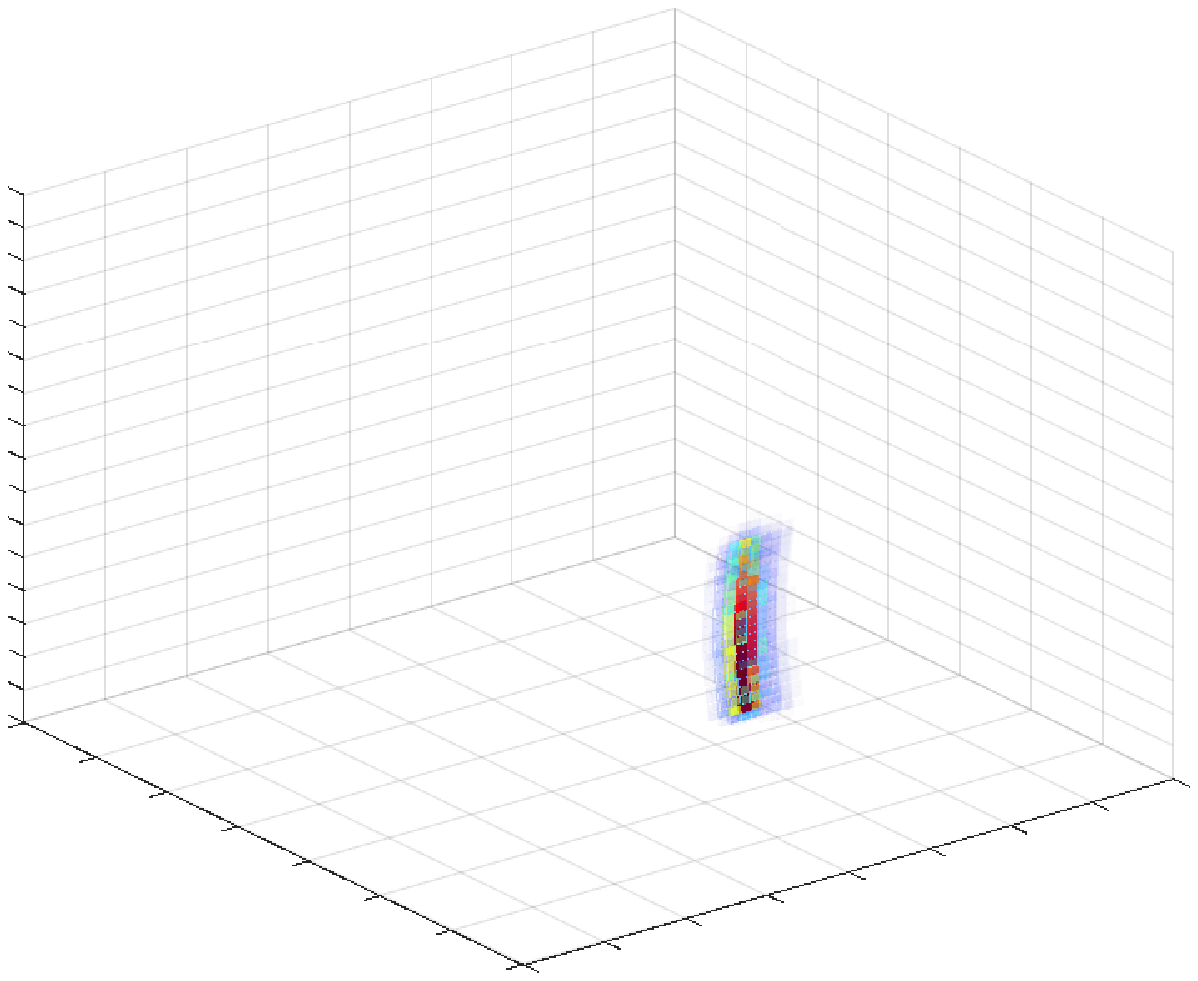}	
			\setlength{\epsfxsize}{1.6in}
			\epsffile{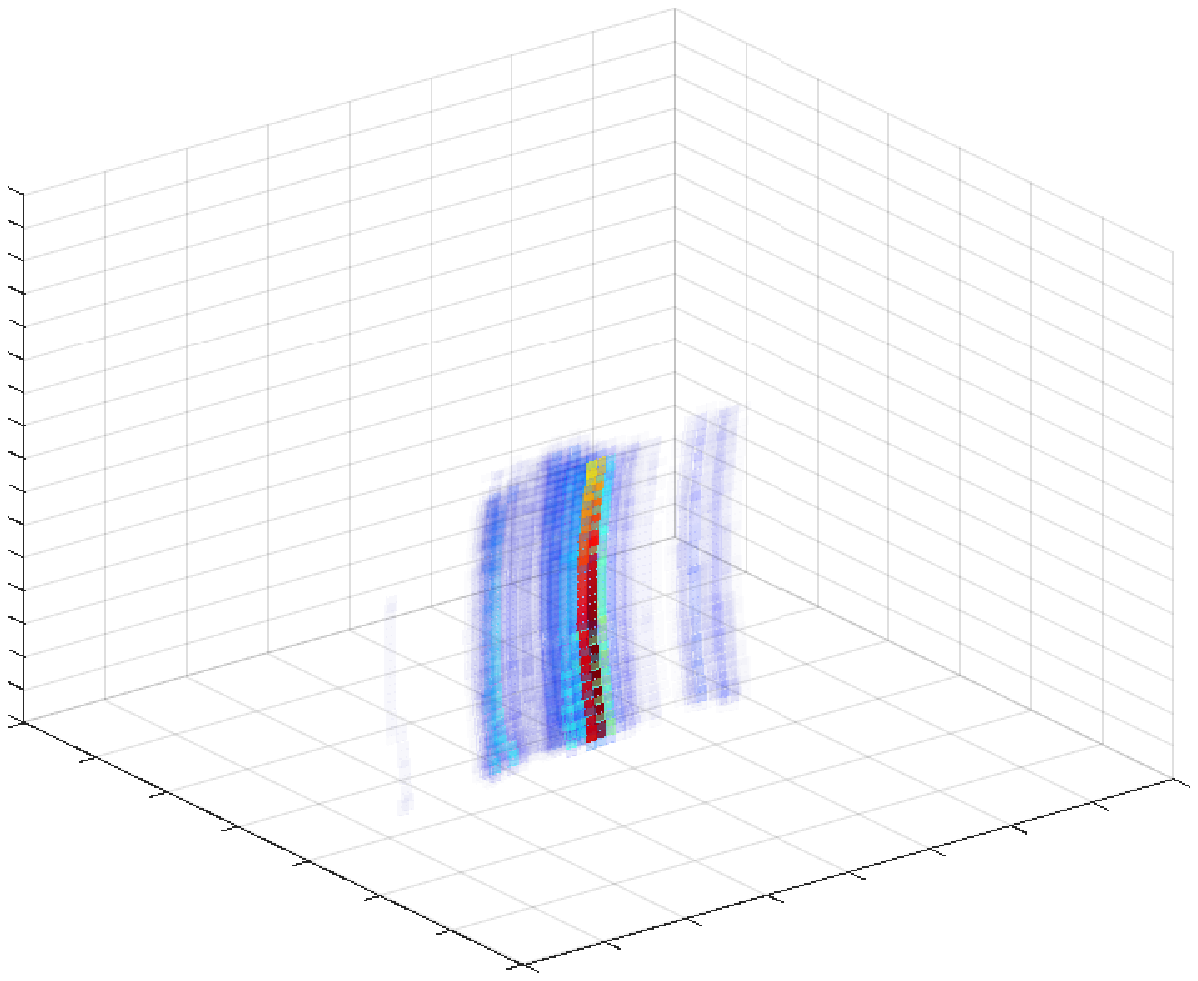}
			\setlength{\epsfxsize}{1.6in}
			\epsffile{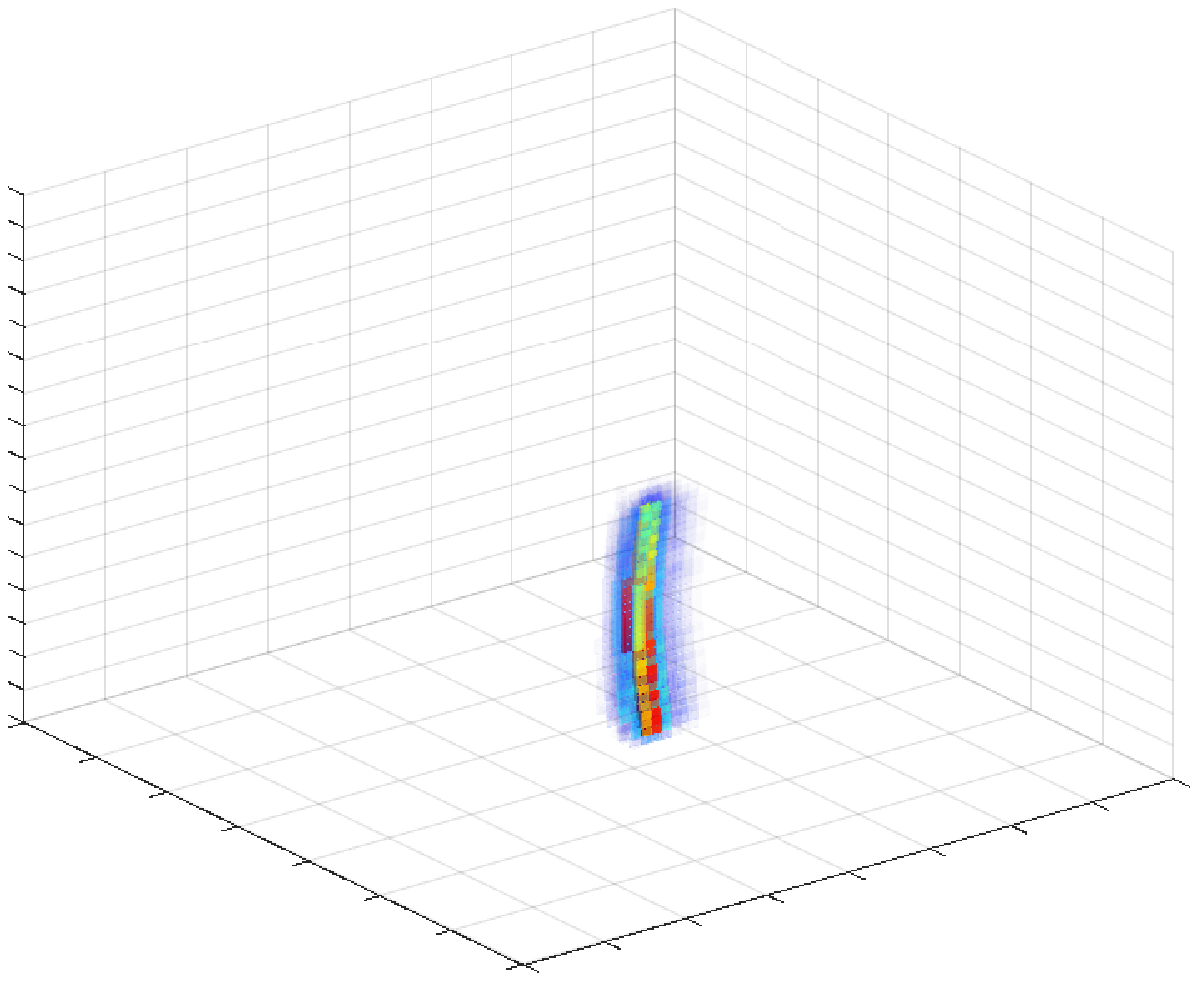}
		\end{center}
	\end{minipage}\label{fig_car6-obj6_radar}\\
	
	\begin{minipage}{7.0in}
		\begin{center}
			\setlength{\epsfxsize}{1.6in}
			\epsffile{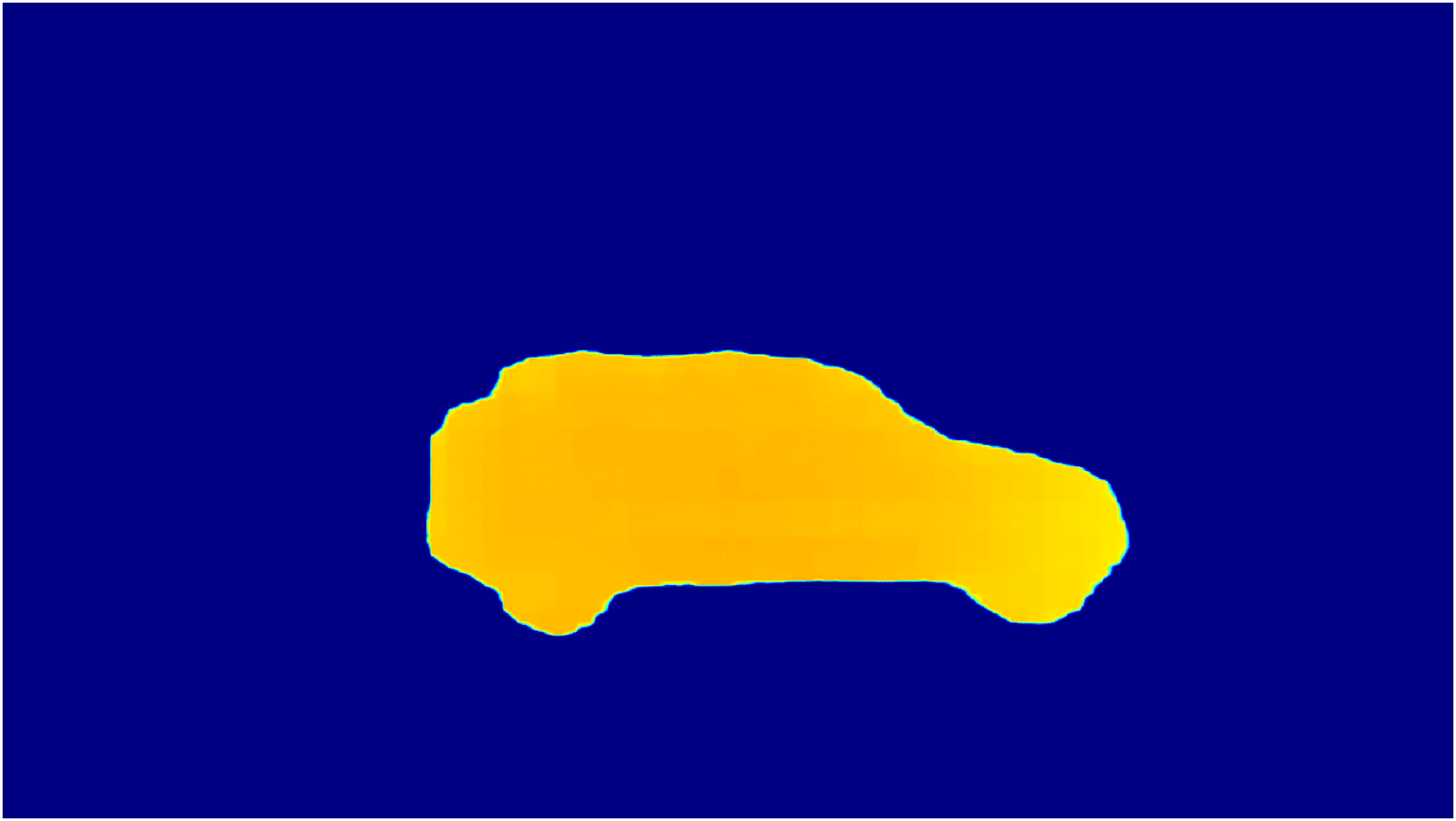}
			\setlength{\epsfxsize}{1.6in}
			\epsffile{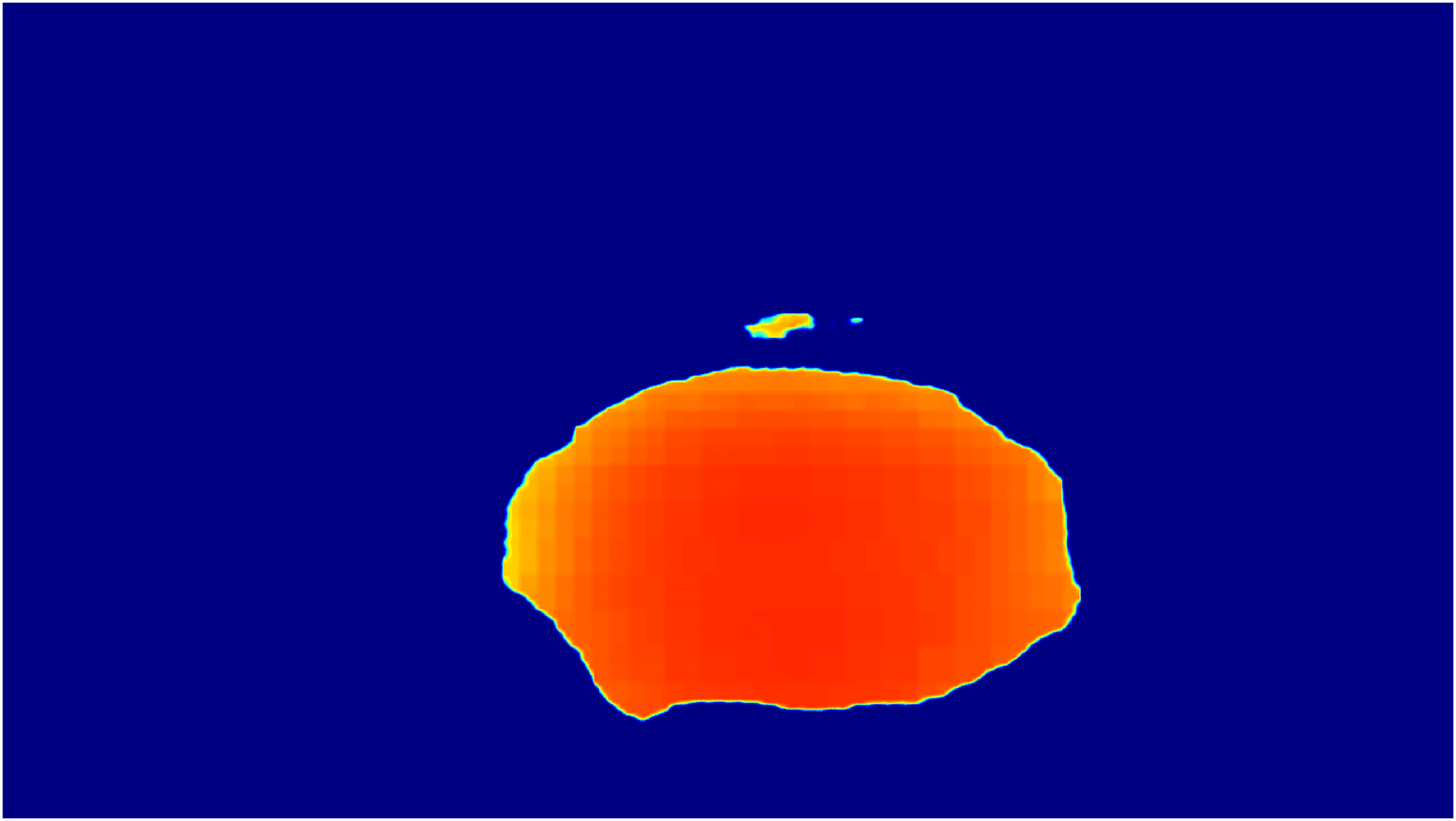}			
			\setlength{\epsfxsize}{1.6in}
			\epsffile{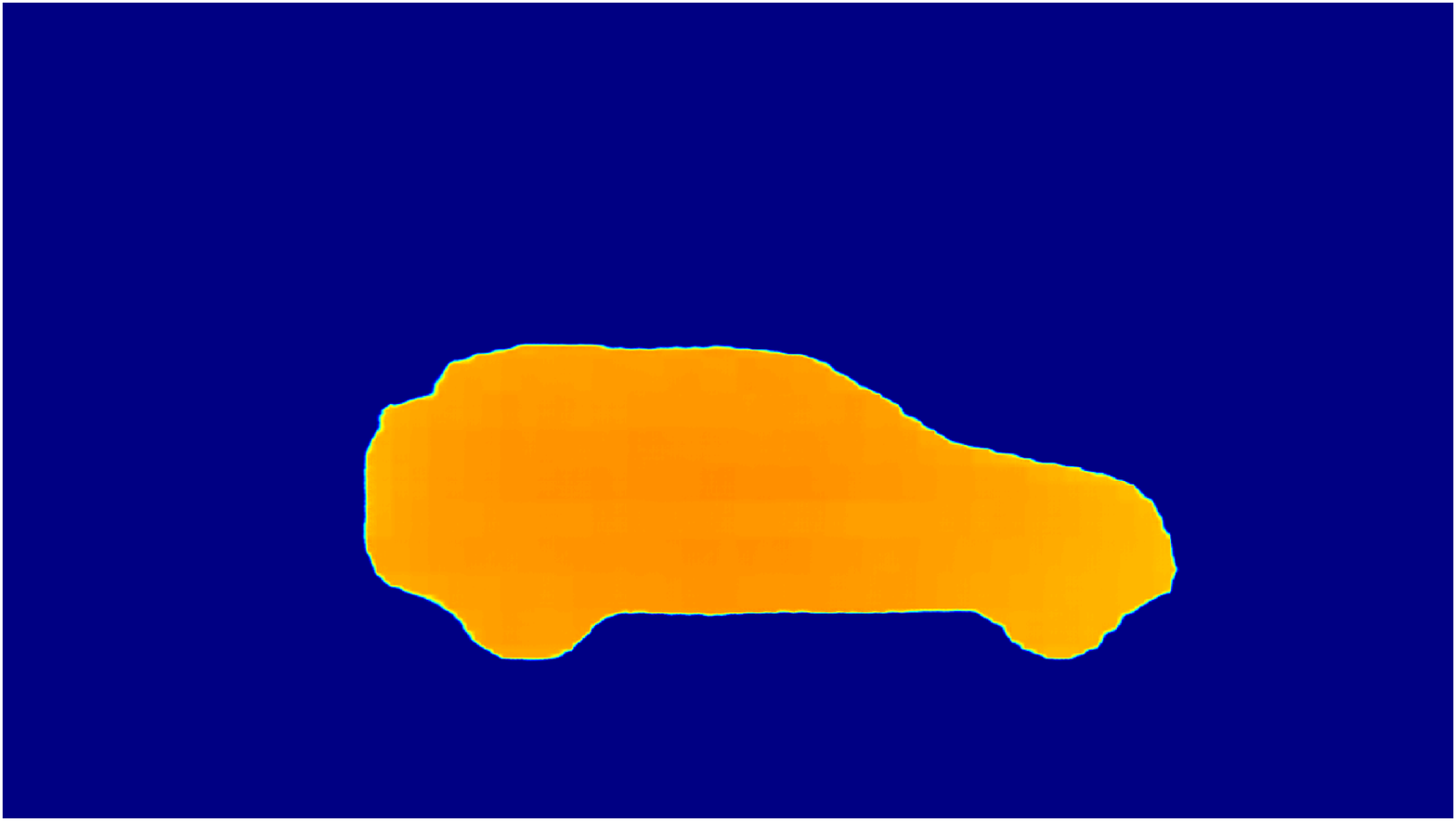}
			\setlength{\epsfxsize}{1.6in}
			\epsffile{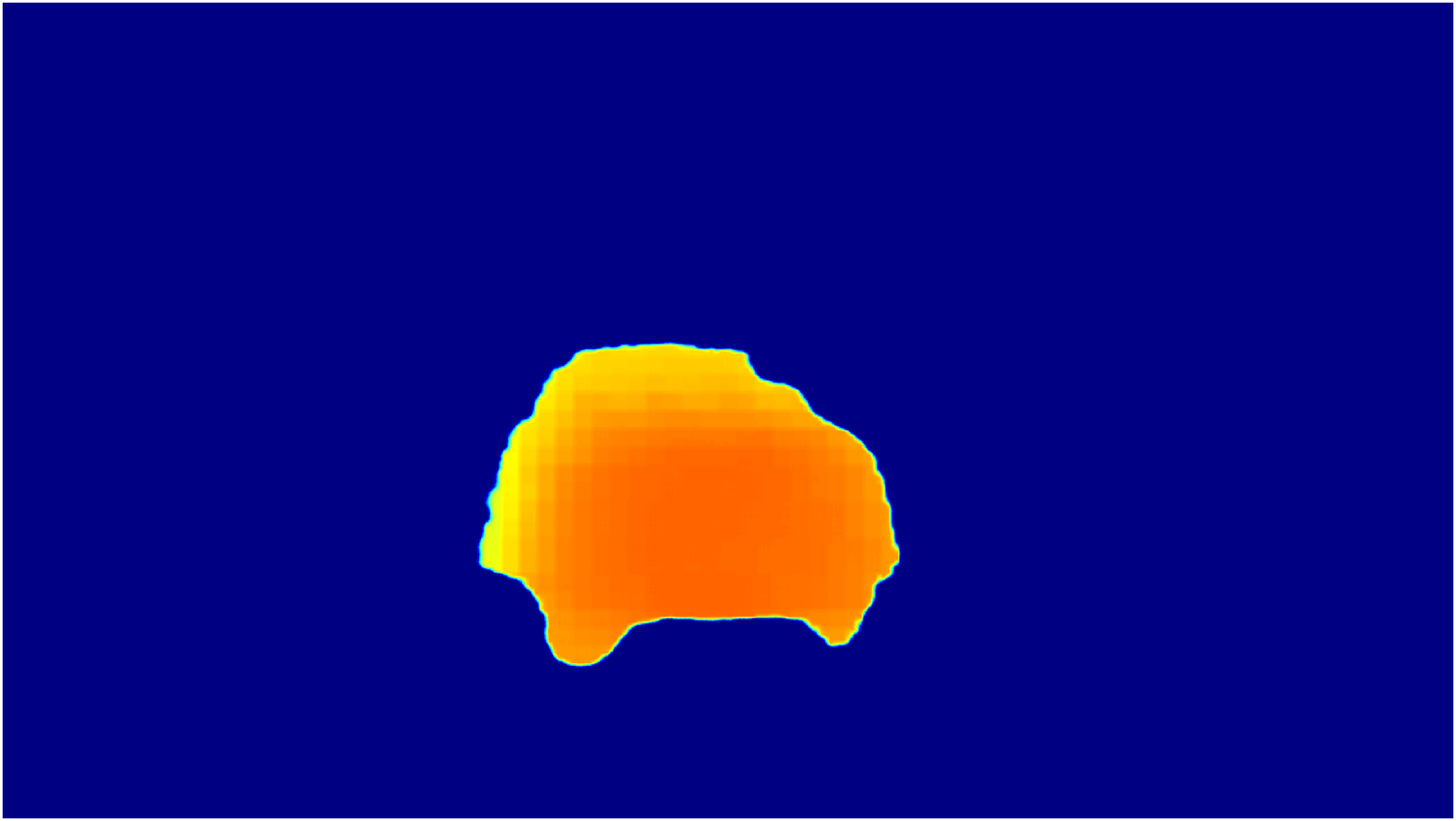}				
		\end{center}
	\end{minipage}\vspace{0.05in}\label{fig_car6-obj6_fake-depth}\\
	\begin{minipage}{7.0in}
		\begin{center}
			\setlength{\epsfxsize}{1.6in}
			\epsffile{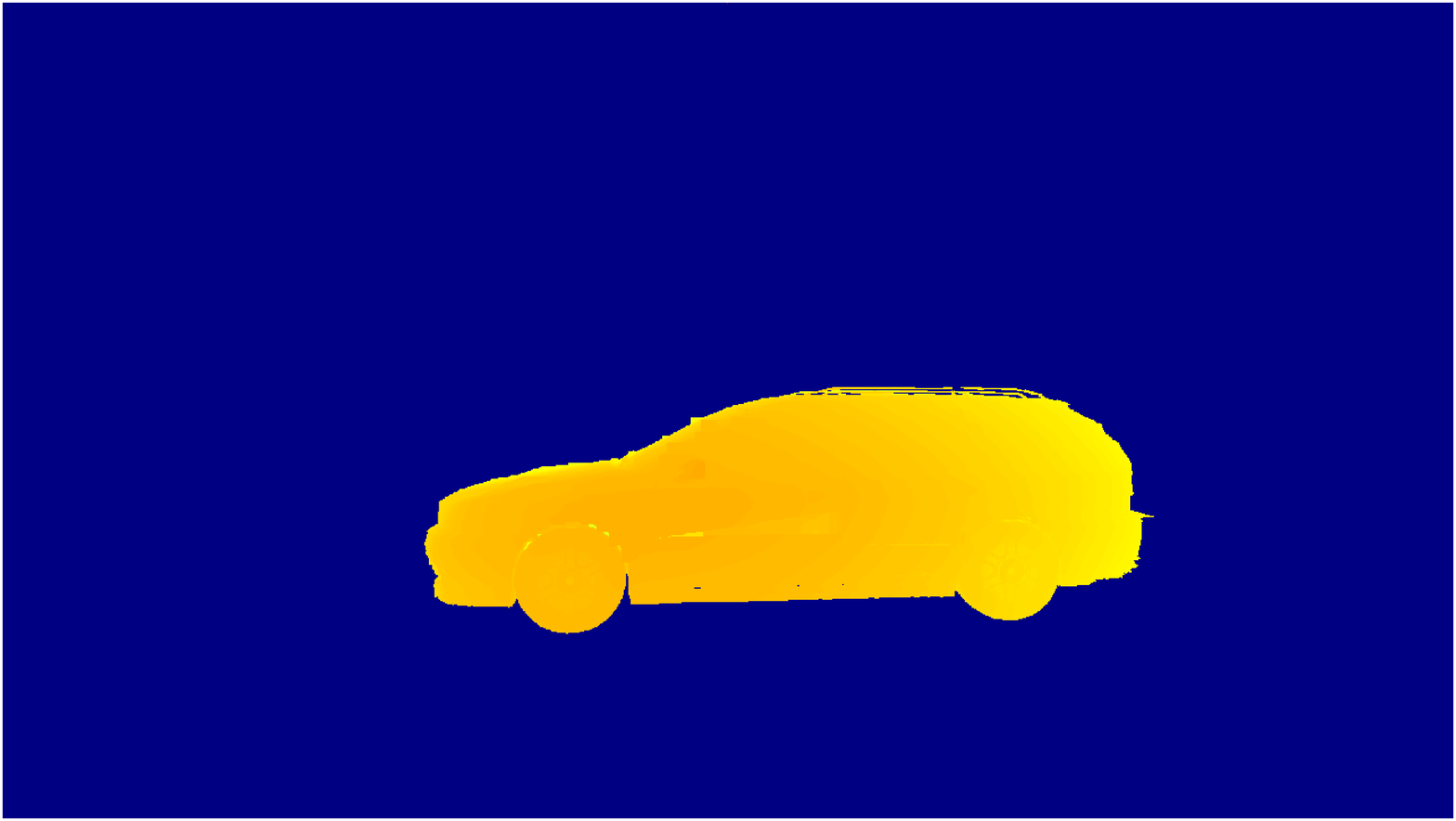}
			\setlength{\epsfxsize}{1.6in}
			\epsffile{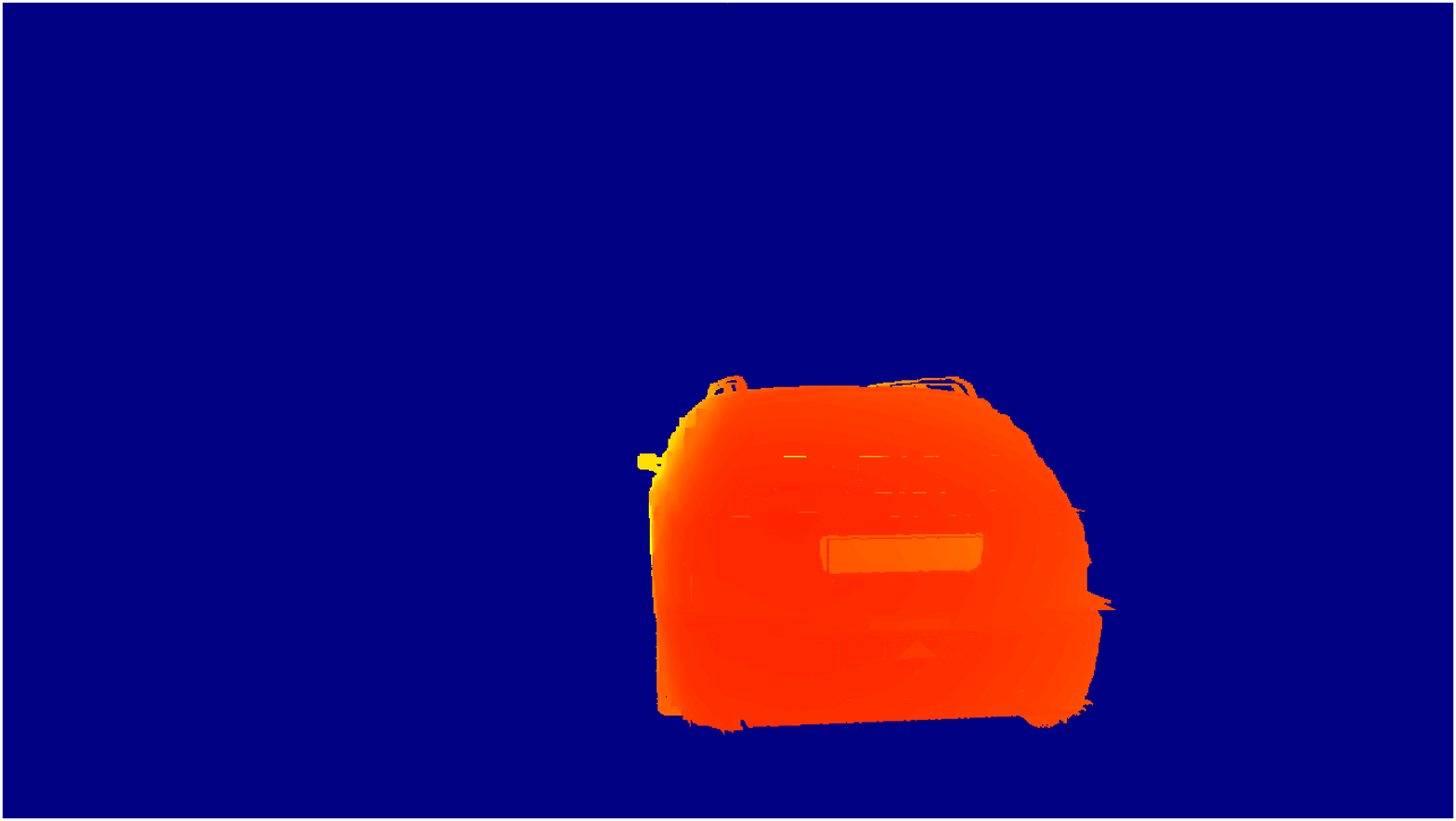}
			\setlength{\epsfxsize}{1.6in}
			\epsffile{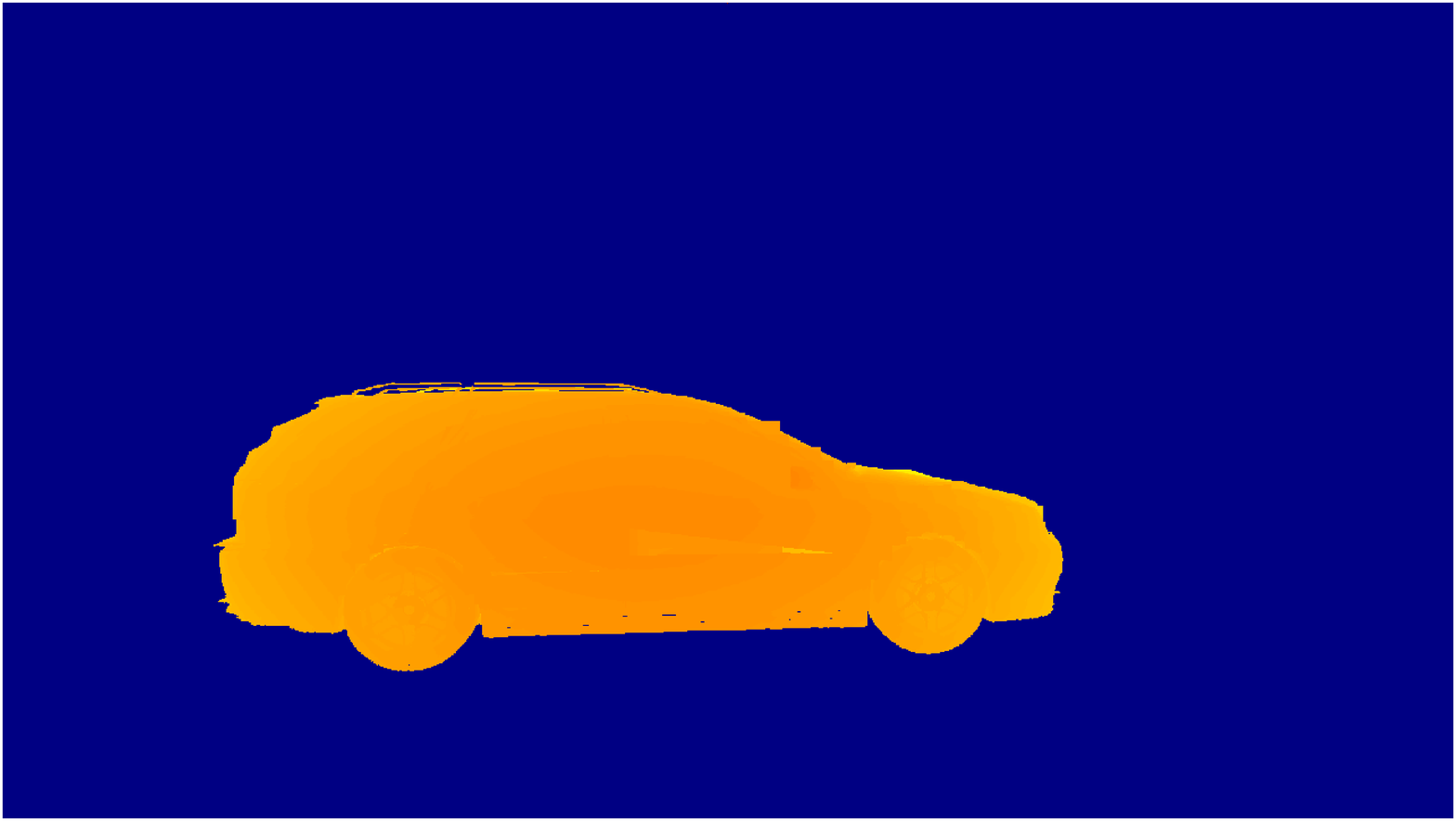}			
			\setlength{\epsfxsize}{1.6in}
			\epsffile{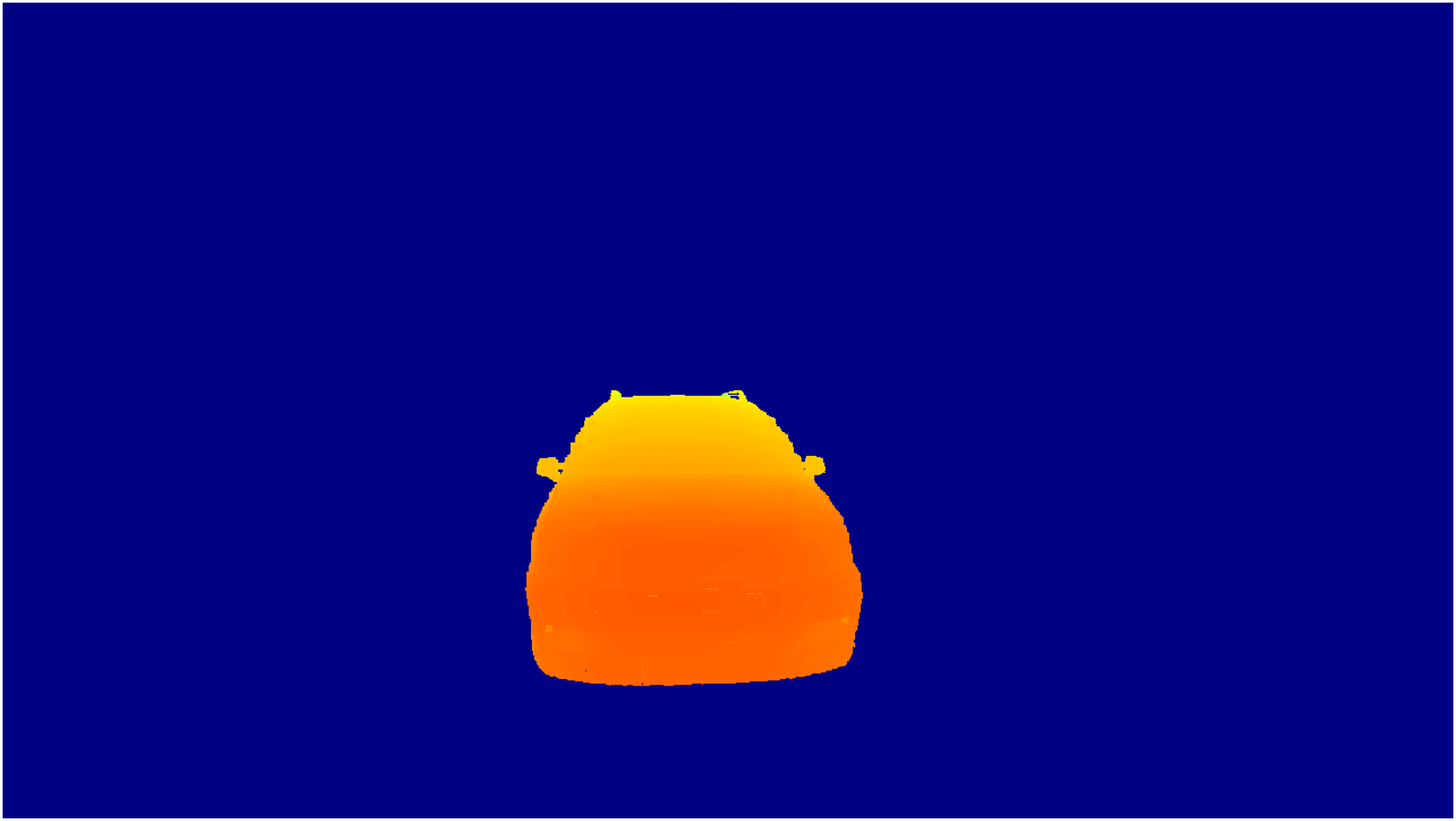}					
		\end{center}
		\label{fig_car6-obj6_real-depth}
	\end{minipage}
	\caption{Example results of generated 2D depth images from the Stage 1 of 3DRIMR. The generated images
		are used as inputs to DeepPoint (which is used as the Stage 2 of 3DRIMR).
	The 1st row shows the 3D radar intensity data from 2 snapshots only. The 2nd row shows the outputs from 3DRIMR's Stage 1. The 3rd row shows the ground truth depth images.}
	\label{fig_stage1_car}
\end{figure*}

We use the 2D depth images generated in the Stage 1 of 3DRIMR to form a dataset of coarse and sparse point clouds,
which includes $1600$ point clouds with $200$ point clouds of each car model. 
We train the proposed generator network and discriminator network 
for $200$ epochs using $1520$ point clouds with batch size $4$.
The learning rate for the first 100 epochs is $2 \times 10^{-4}$ and linearly decreases to 0 in the rest 100 epochs.
Then we test the generator network using the remaining $80$ point clouds.
Fig. \ref{fig_stage1_car} shows some example results of generated 2D depth images from the Stage 1 of 3DRIMR. 

\begin{figure*}[htb!]
	\begin{minipage}{7.2in}
		\begin{center}
			\setlength{\epsfxsize}{1.7in}
			\epsffile{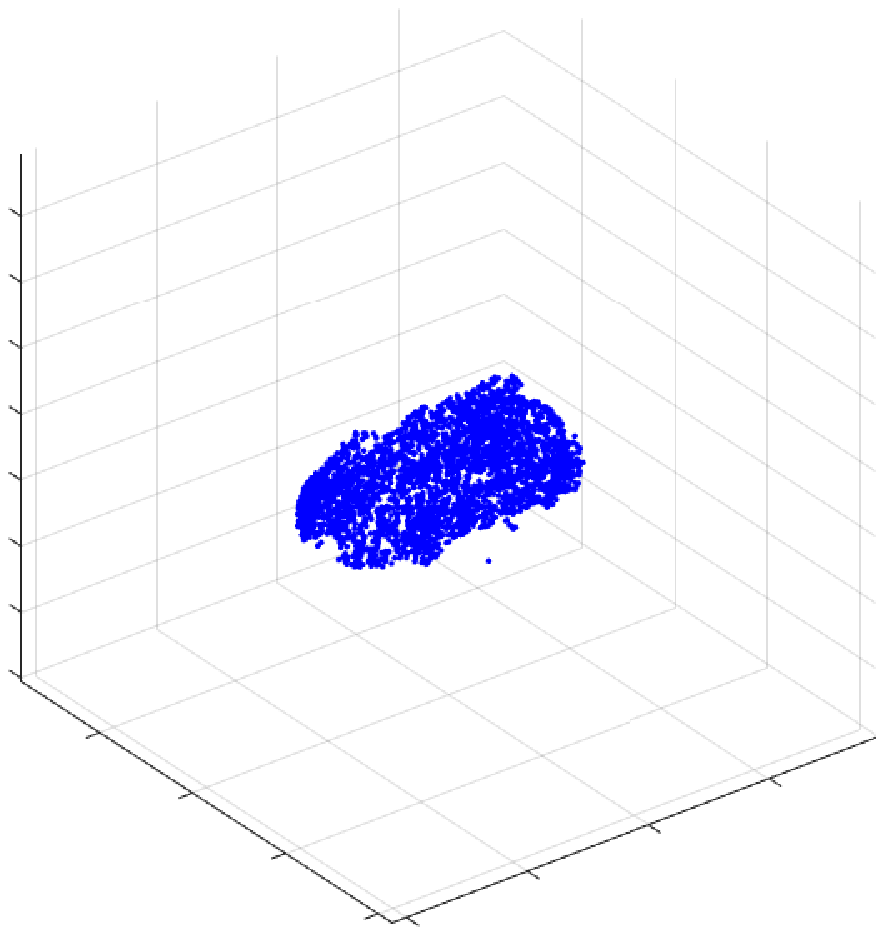}
			\setlength{\epsfxsize}{1.7in}
			\epsffile{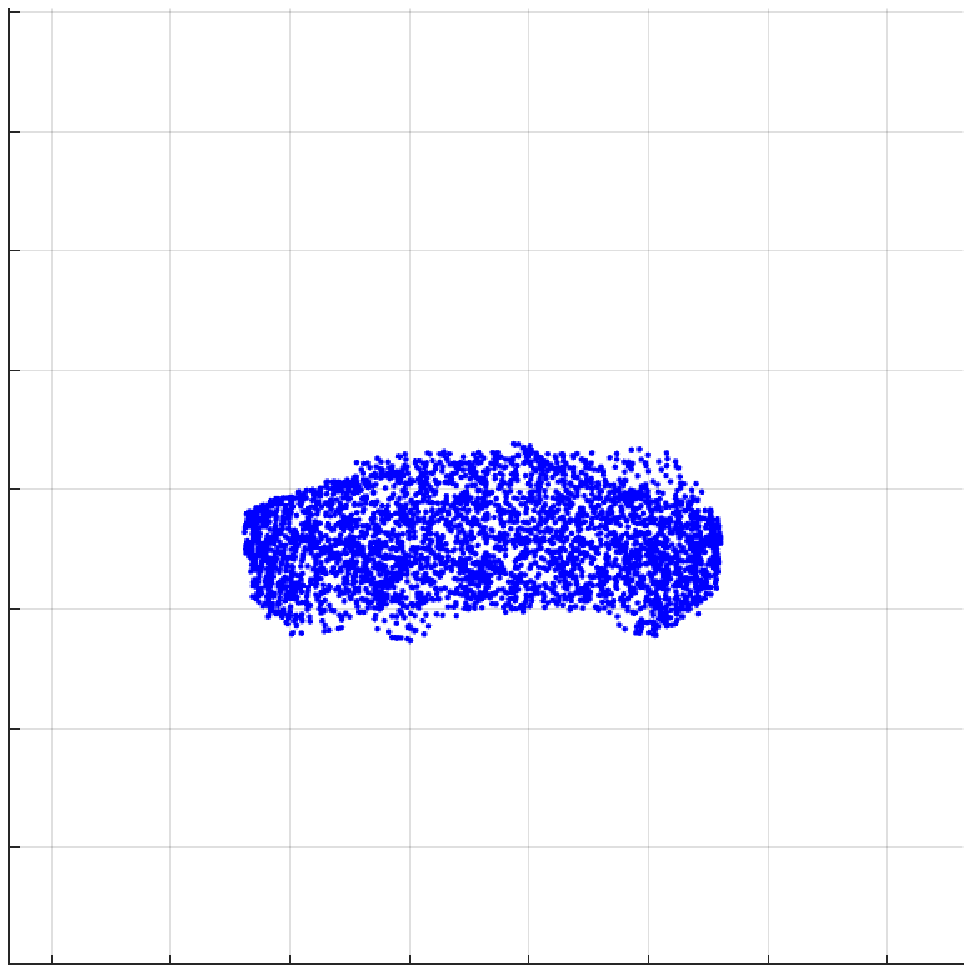}	
			\setlength{\epsfxsize}{1.7in}
			\epsffile{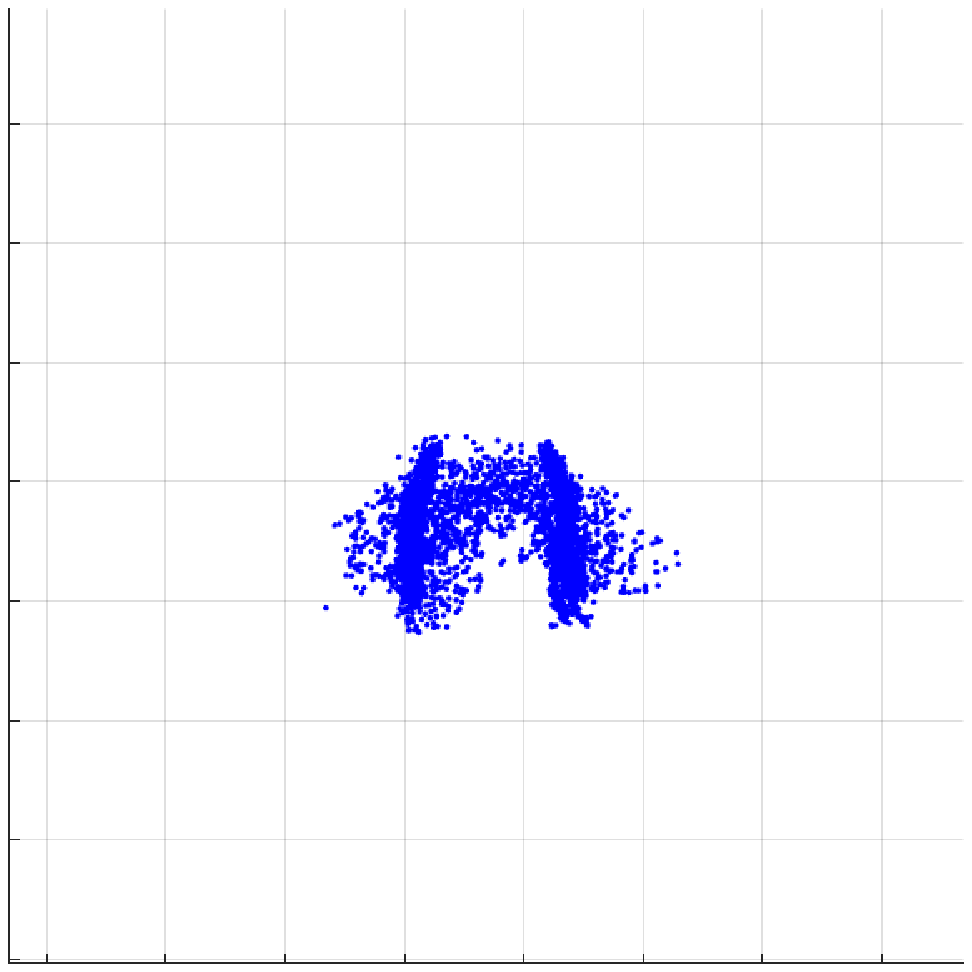}
			\setlength{\epsfxsize}{1.7in}
			\epsffile{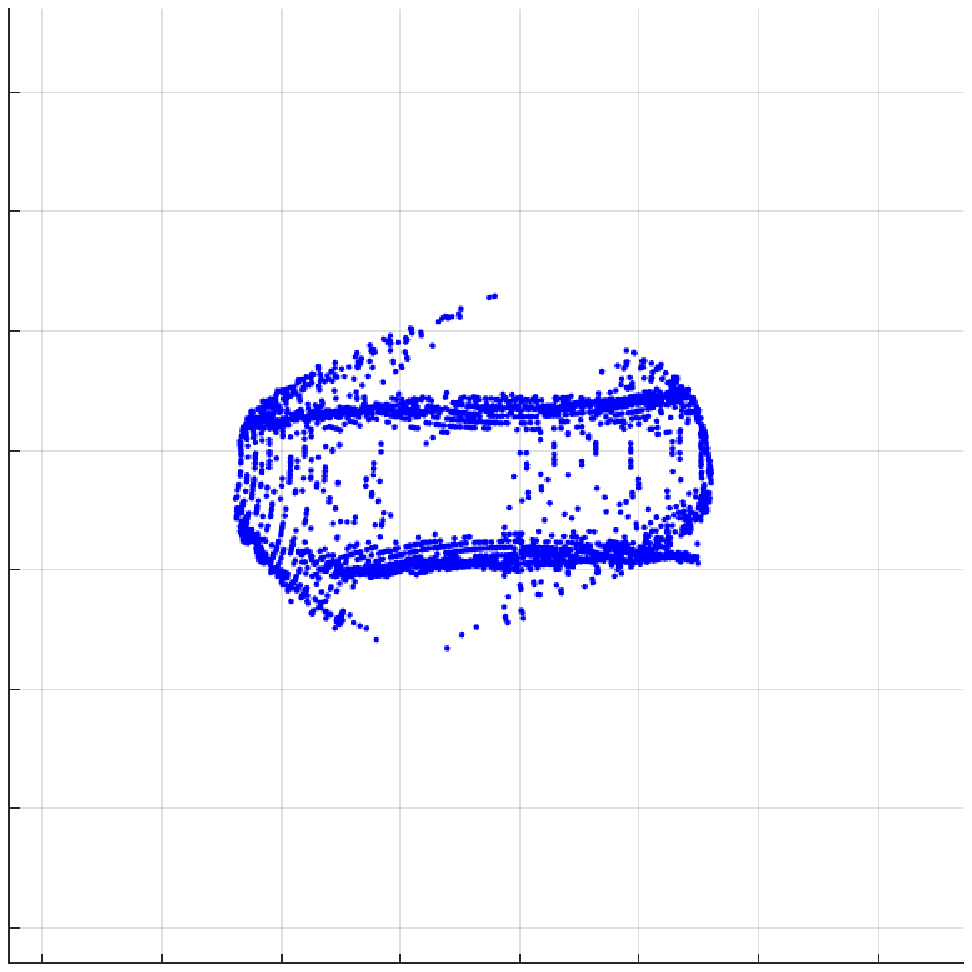}
		\end{center}
	\end{minipage}\label{fig_car6-obj6-fake-pc}\\
    \begin{minipage}{7.2in}
    	\begin{center}
    		\setlength{\epsfxsize}{1.7in}
    		\epsffile{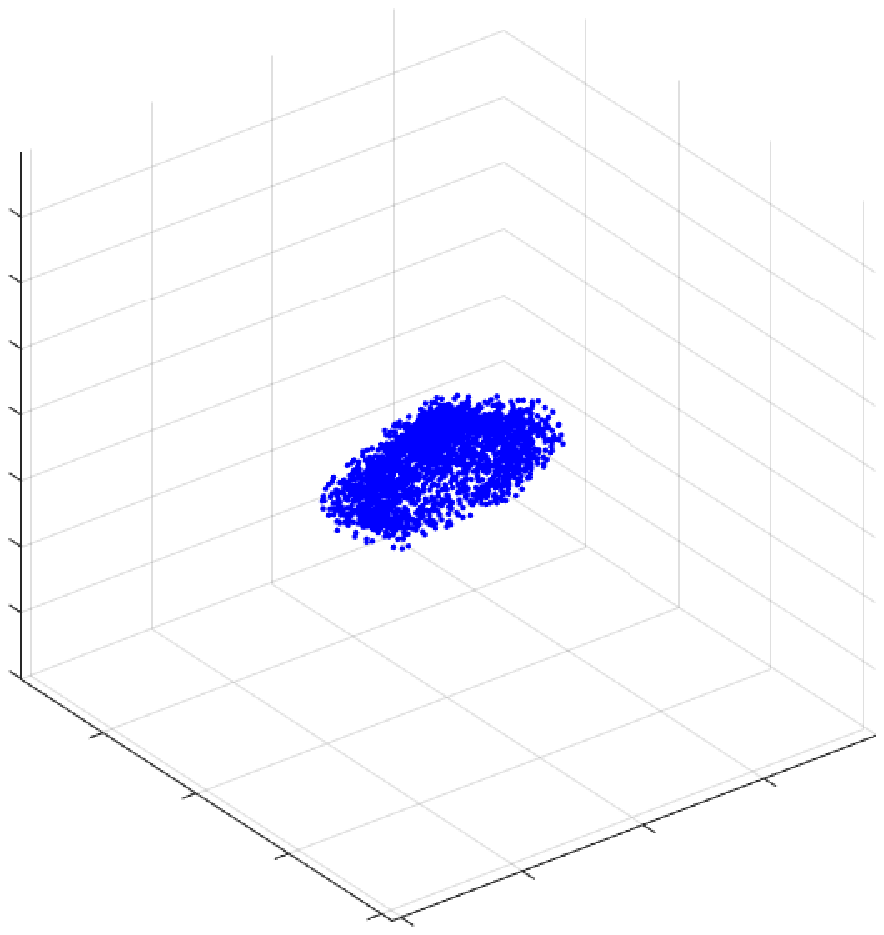}
    		\setlength{\epsfxsize}{1.7in}
    		\epsffile{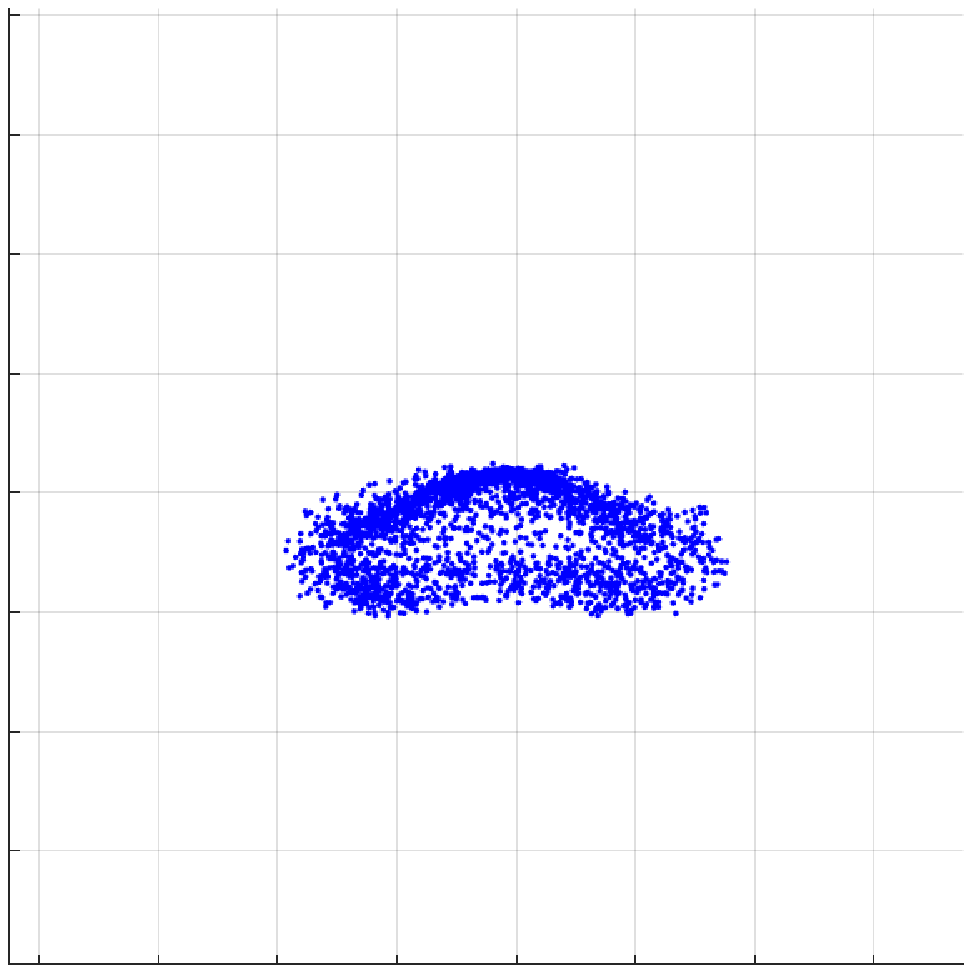}	
    		\setlength{\epsfxsize}{1.7in}
    		\epsffile{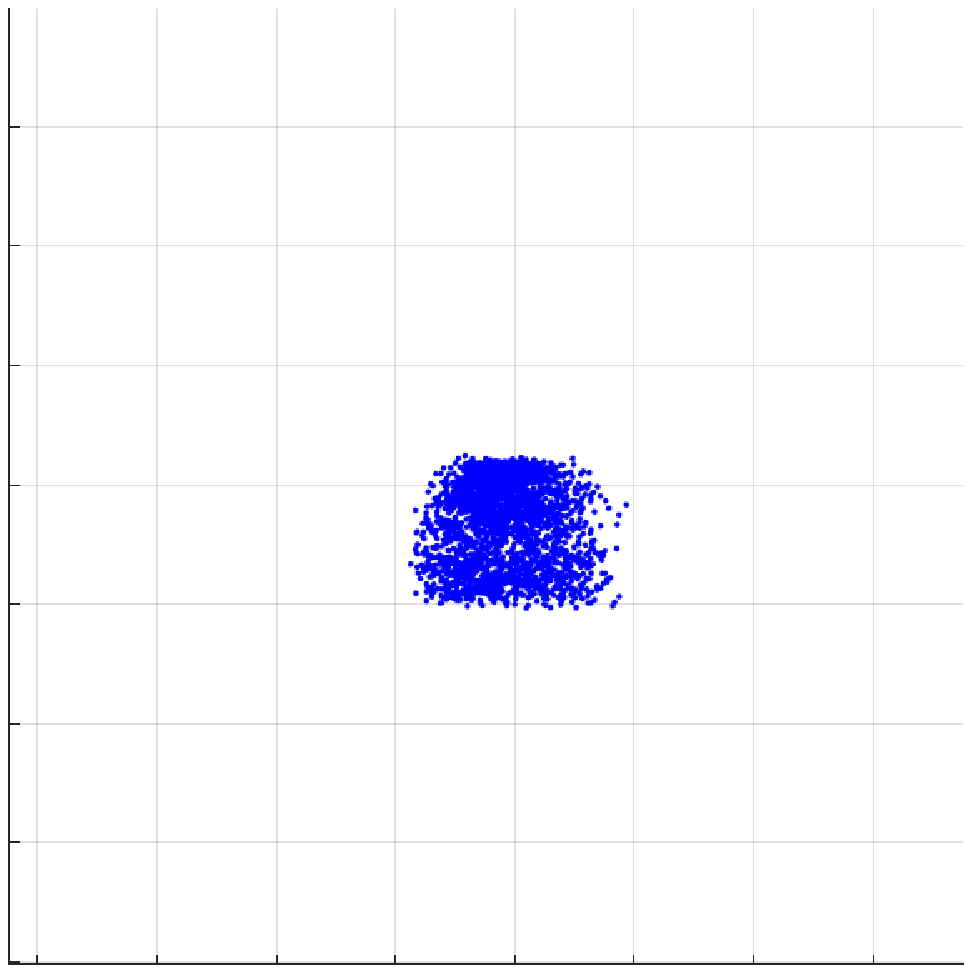}
    		\setlength{\epsfxsize}{1.7in}
    		\epsffile{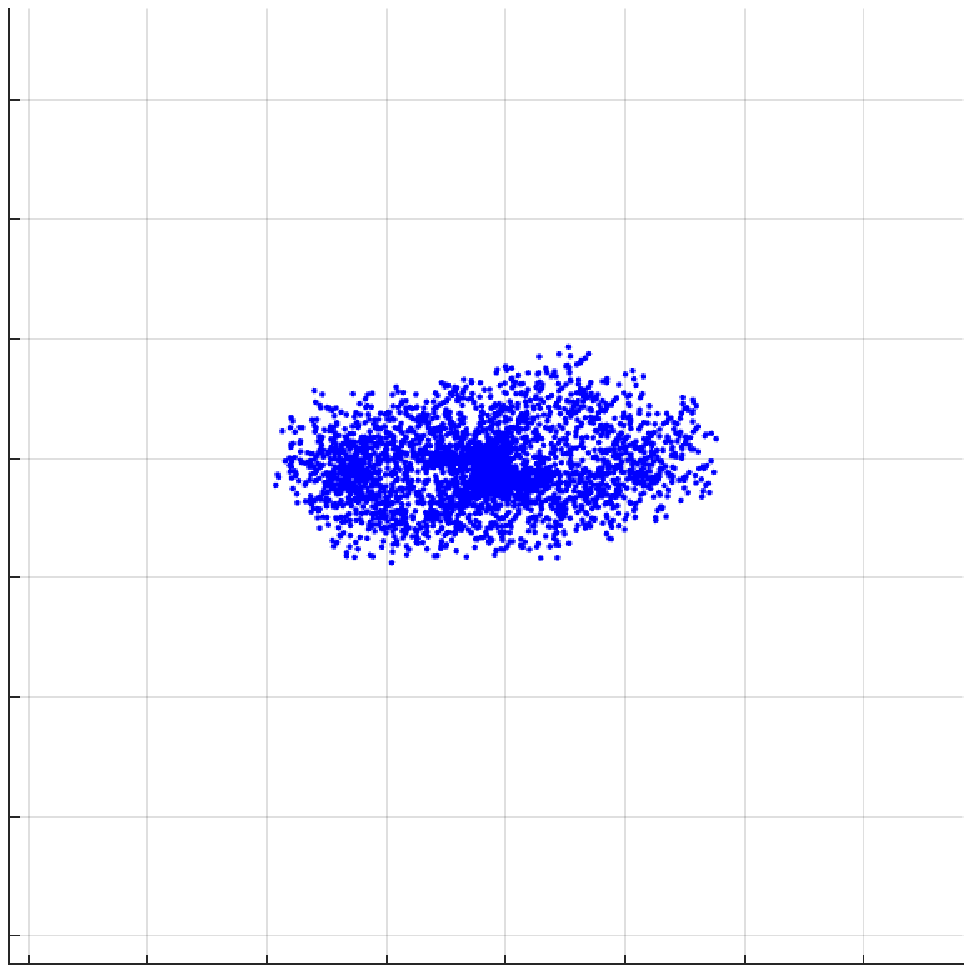}
    	\end{center}
    \end{minipage}\vspace{0.05in}\label{fig_car6-obj6-out-pc}\\	
	\begin{minipage}{7.2in}
		\begin{center}
			\setlength{\epsfxsize}{1.7in}
			\epsffile{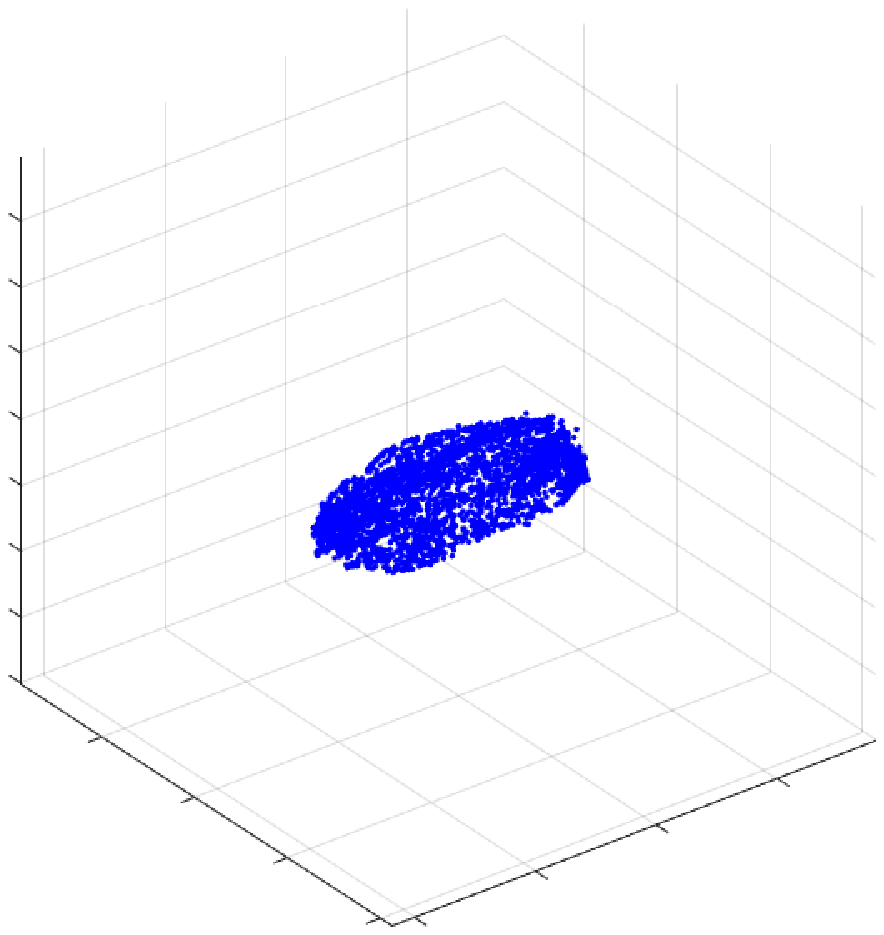}
			\setlength{\epsfxsize}{1.7in}
			\epsffile{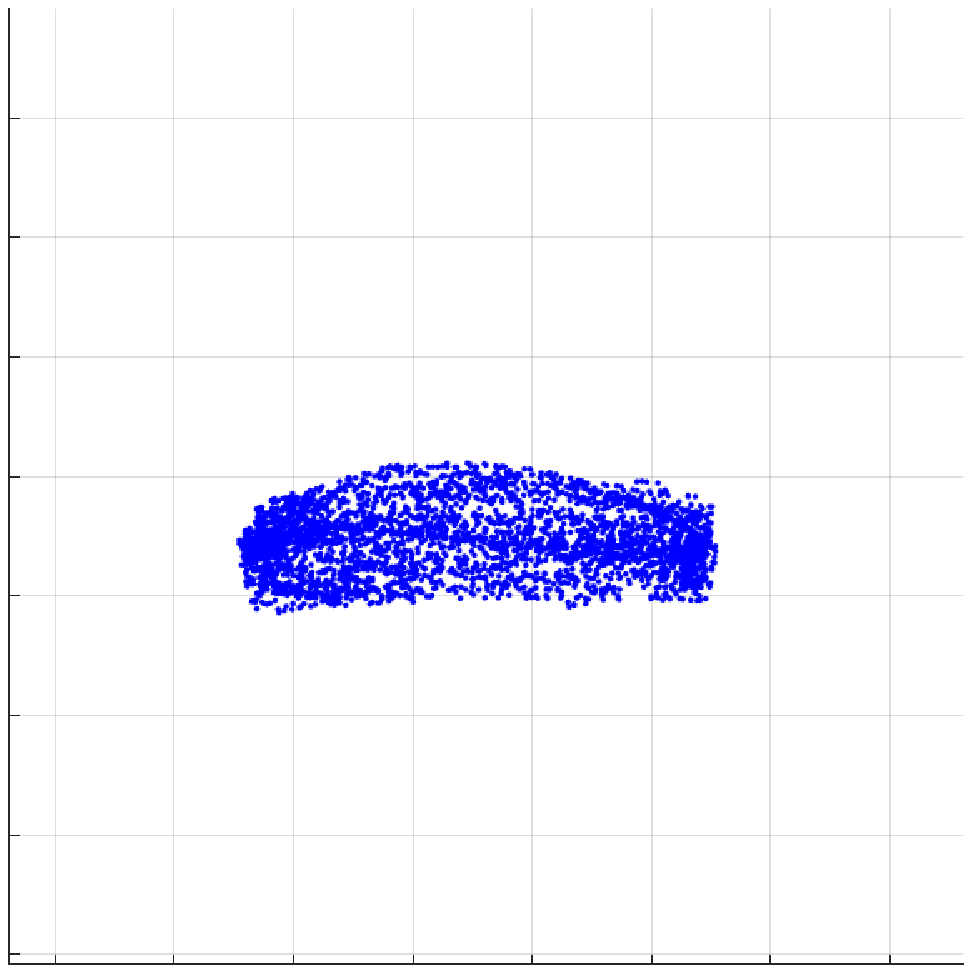}	
			\setlength{\epsfxsize}{1.7in}
			\epsffile{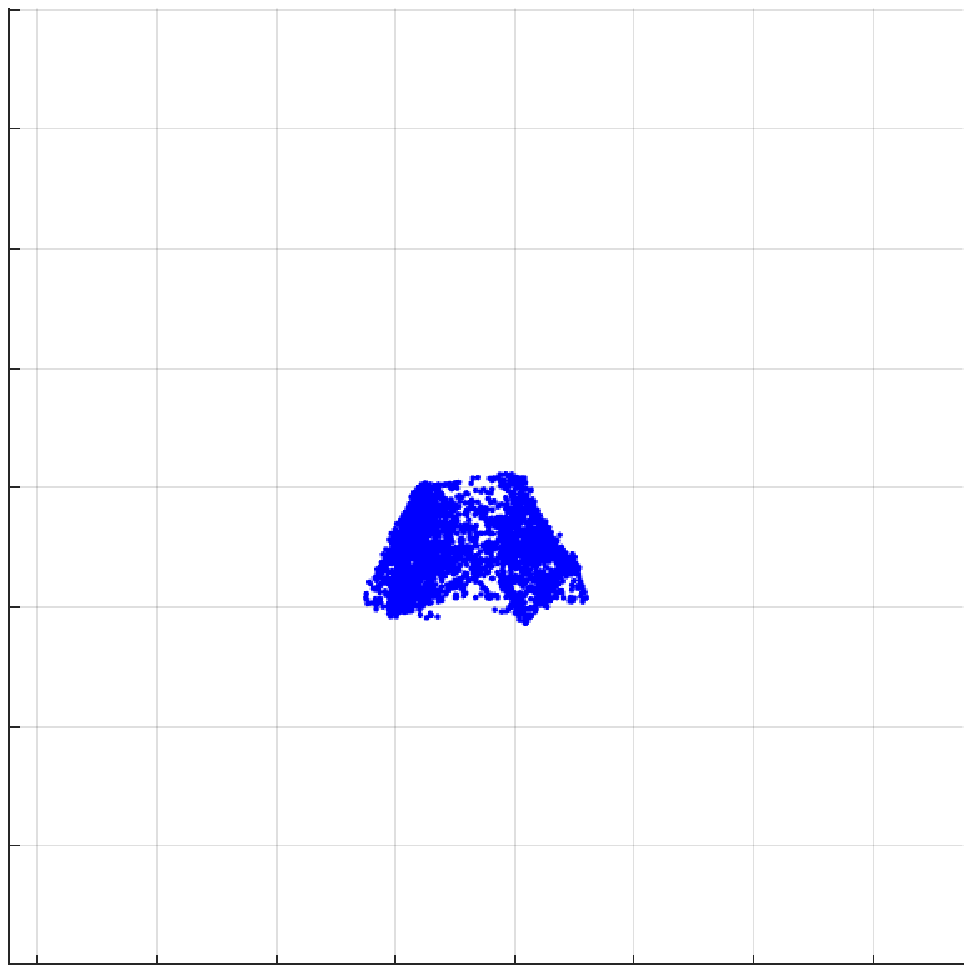}
			\setlength{\epsfxsize}{1.7in}
			\epsffile{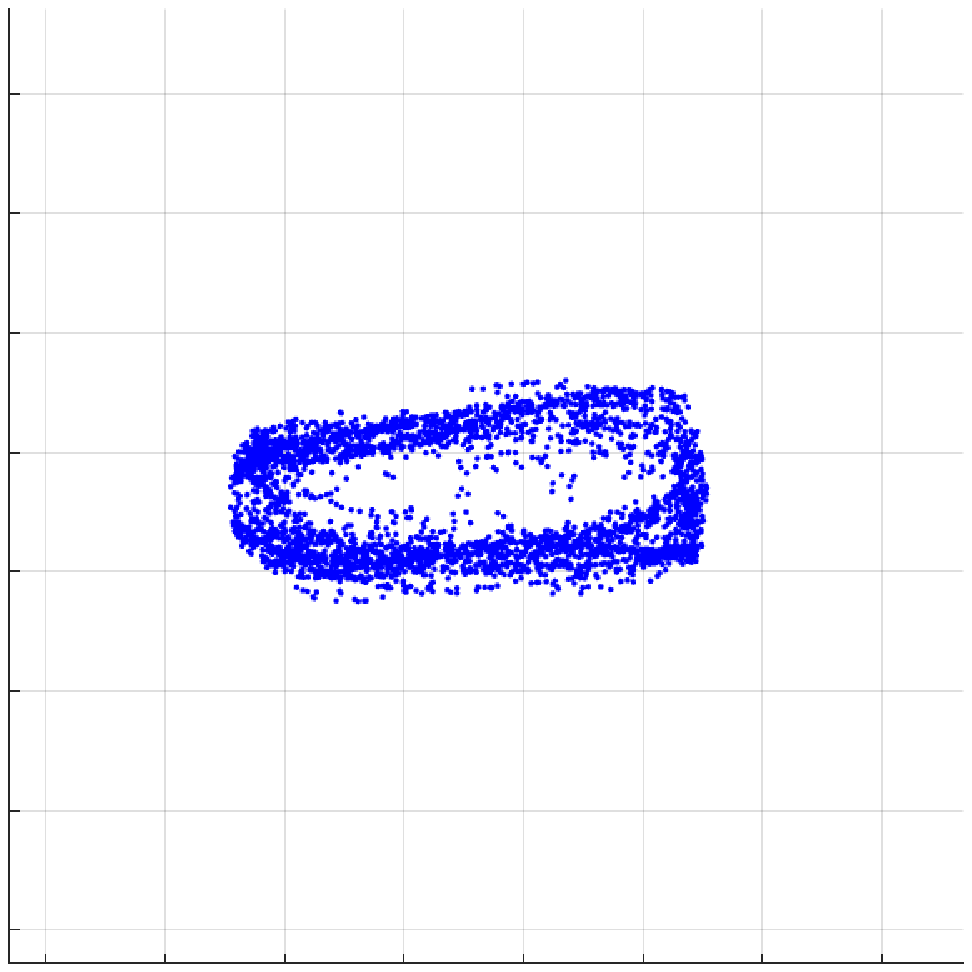}
		\end{center}
	\end{minipage}\vspace{0.05in}\label{fig_car6-obj6-out-pc}\\
	\begin{minipage}{7.2in}
		\begin{center}
			\setlength{\epsfxsize}{1.7in}
			\epsffile{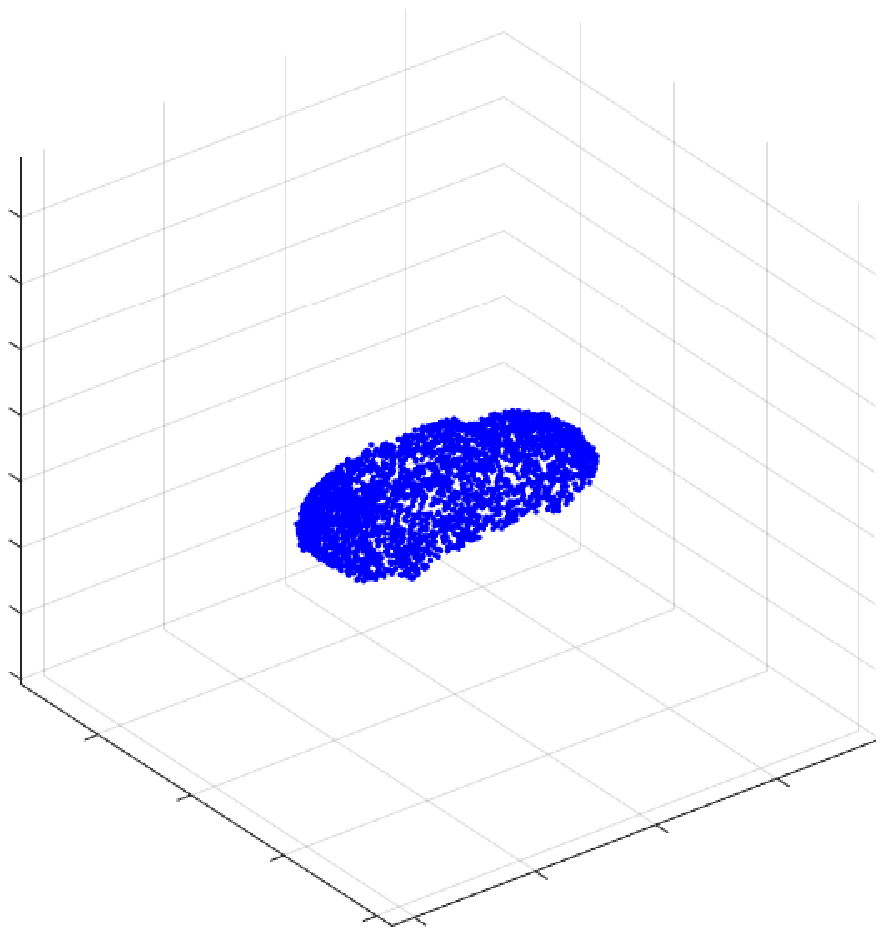}
			\setlength{\epsfxsize}{1.7in}
			\epsffile{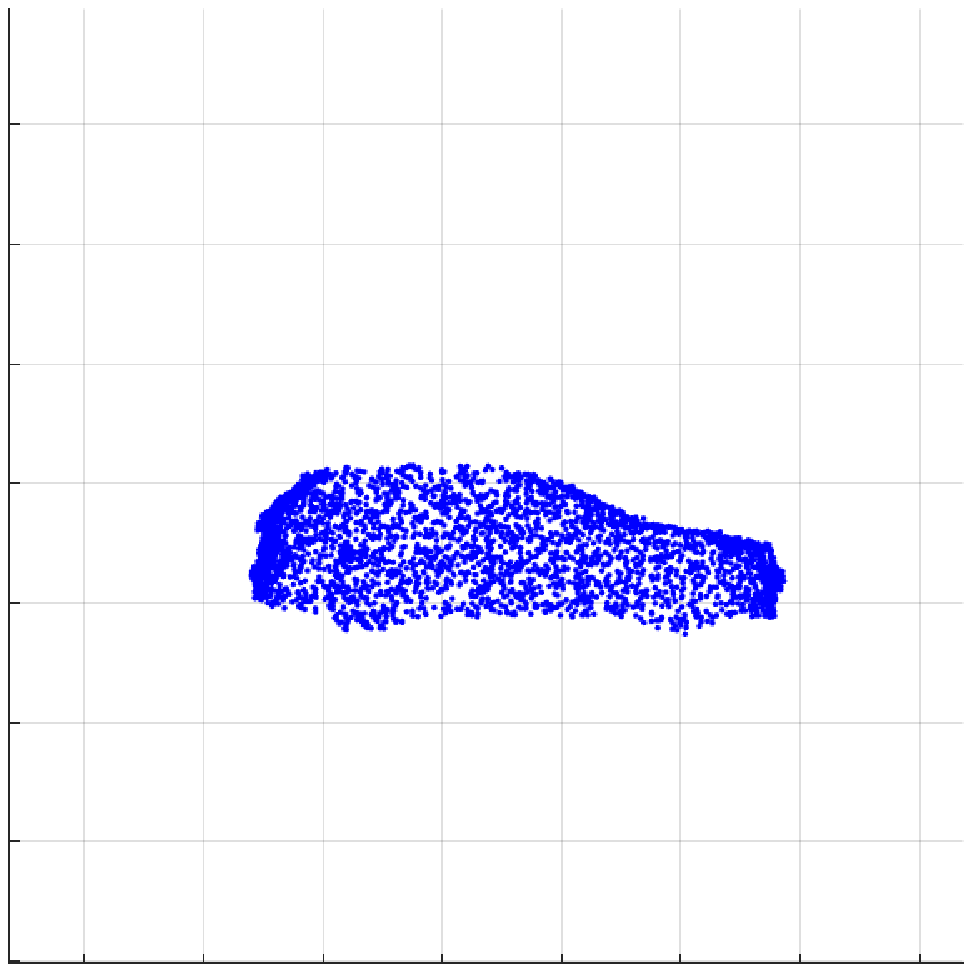}	
			\setlength{\epsfxsize}{1.7in}
			\epsffile{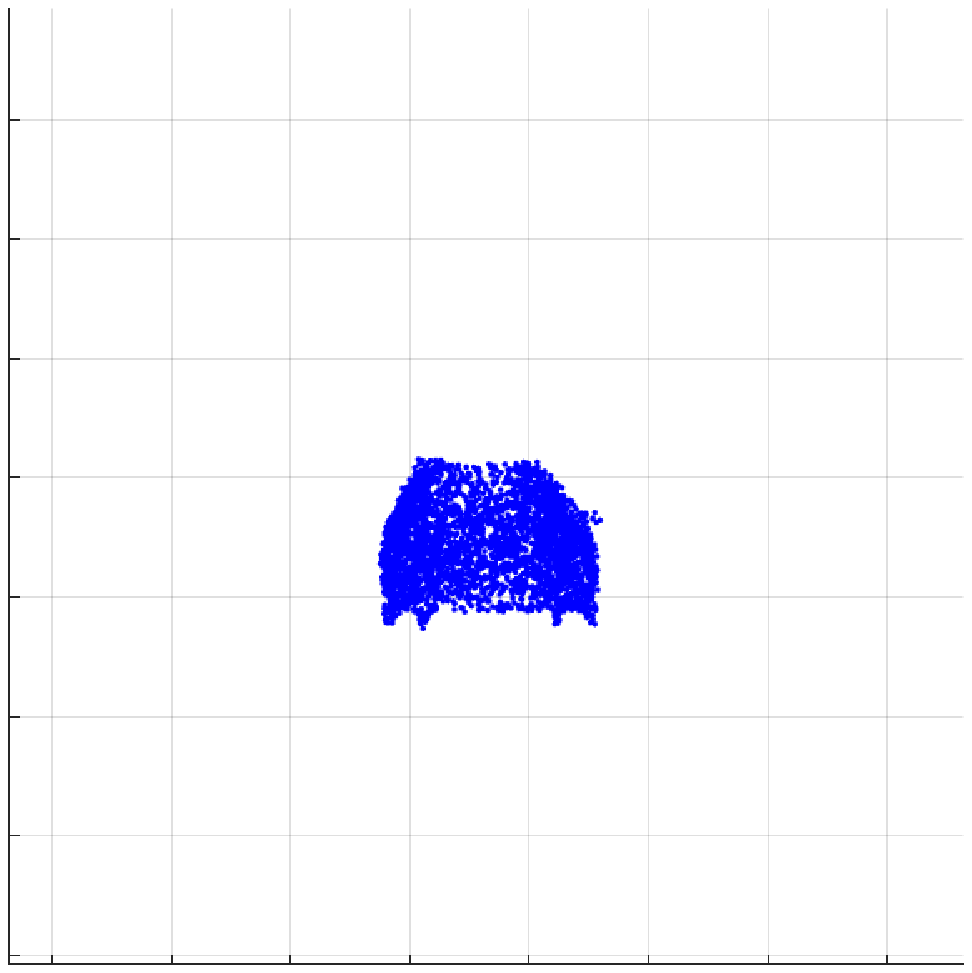}
			\setlength{\epsfxsize}{1.7in}
			\epsffile{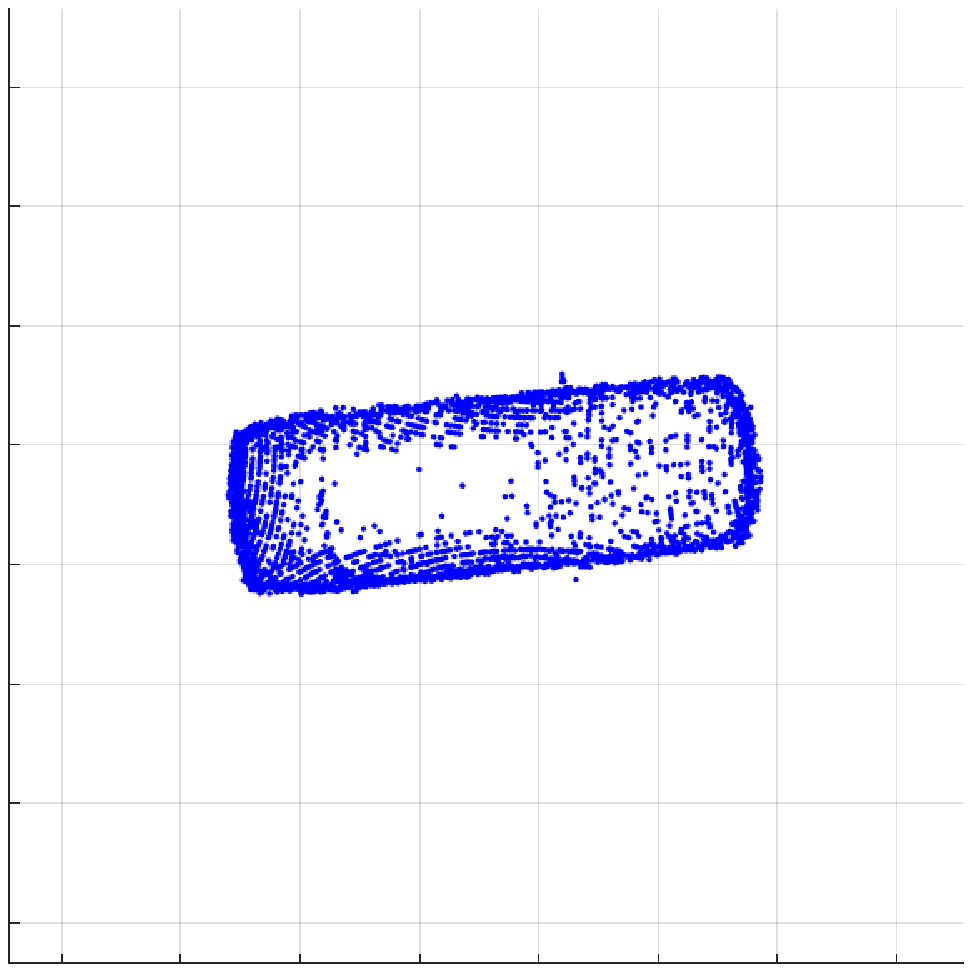}
		\end{center}
	\end{minipage}\vspace{0.05in}\label{fig_car6-obj6-real-pc}\\	
	\caption{Example experimental results of the proposed generator network. The 1st row shows the input point clouds of a car from different viewpoints (i.e., inputs to the generator network). The 2nd row shows the output point clouds of 3DRIMR. The 3rd row shows the output point clouds of DeepPoint. The 4th row shows the ground truth point clouds. Counting from the side, the 1st column lists point clouds shown in 3D space. The 2nd column lists the front views of point clouds. The 3rd column shows the side views of point clouds. The 4th column shows the top views of point clouds. }
	\label{fig_stage2_car_pc}
\end{figure*}


\subsection{Evaluation Results}

In this section, we compare our generator network performance with 
the original 3DRIMR \cite{sun20213drimr} 
since both of them aim at reconstructing objects' 3D point clouds from sparse radar data.
We also validate our design choices for generator and discriminator
networks by controlled experiments.
We conduct experiments by varying the number of layers, i.e., DeepPoint blocks, 
in our proposed generator network 
model. Specifically, as shown in Table \ref{tab_stage2_results_all}, we tested $1, 2, 5$ and $7$ layers of DeepPoint blocks
in our generator network. 

\renewcommand{\arraystretch}{1.5} 
\begin{table}[tp]	
	\centering
	\fontsize{6.5}{8}\scriptsize
	\begin{tabular}{|c|c|c|c|c|c|c|}
		\hline
		\multirow{2}{*}{Method}&
		\multicolumn{2}{c|}{CD}&\multicolumn{2}{c|}{EMD}&\multicolumn{2}{c|}{ F-score}\cr\cline{2-7}
		&avg.&std.&avg.&std.&avg.&std.\cr
		\hline
		\hline
		3DRIMR                                     & 7.89       &4.11            & -              & -                 & 8.41            & 3.22      \cr\hline
		1-Block, w/o sc                          & 10.10     &4.49           & 5.01         & 4.24           & 8.40           & 3.40        \cr\hline
		1-Block                                      & 9.75      &4.00           &4.56          &3.96            &8.47             &3.44         \cr\hline
		2-Block                                     & 9.40       &4.70           &4.83          &4.84            &9.40             &4.22         \cr\hline
		5-Block                                    & 7.79        &4.37           & {\bf 4.40 } & 4.49           & 13.10           & 5.97          \cr\hline
		7-Block                                    & {\bf 7.68} & 4.15          & 4.53          &  4.19          &{\bf 13.23}    & 6.34        \cr\hline
		7-Block + 3sc                          & 9.13        & 4.55           & 4.66          & 3.88          &  10.70          & 4.90          \cr\hline
	\end{tabular}
	\captionsetup{font={scriptsize}}
	\caption{Quantitative results under different setups. Note that the units of CD and EMD in this table are cm, and the magnitude of F-score is $10^{-2}$.}
	\label{tab_stage2_results_all}
\end{table}

\subsubsection{Comparison with 3DRIMR}

In Table \ref{tab_stage2_results_all}, we can see the our generator network with 5-Block and 7-Block 
significantly outperforms the original 3DRIMR in terms of both Chamfer Distance (CD) and F-score.
Note that EMD results are not available for 3DRIMR. 
In addition, 
Fig. \ref{fig_stage2_car_pc} demonstrates that the visual improvement of output point clouds is even more obvious.
This is because 3DRIMR can only reconstruct an overall shape of the object whereas our proposed
generator network can recover more fine details of the shape, e.g., correct orientation, 
and the shapes of wheels of the car.
This significant improvement is due to the ``deeper" structure of the generator network,
the optimal number of DeepPoint blocks, the introduction of skip connections,  
and the use of a more efficient training loss metric, i.e., Earth Mover's Distance (EMD).

\subsubsection{Performances of different layers of DeepPoint Blocks}
\begin{figure*}[htb!]
	\centerline{
		\begin{minipage}{7.2in}
			\begin{center}
				\setlength{\epsfxsize}{2.2in}
				\epsffile{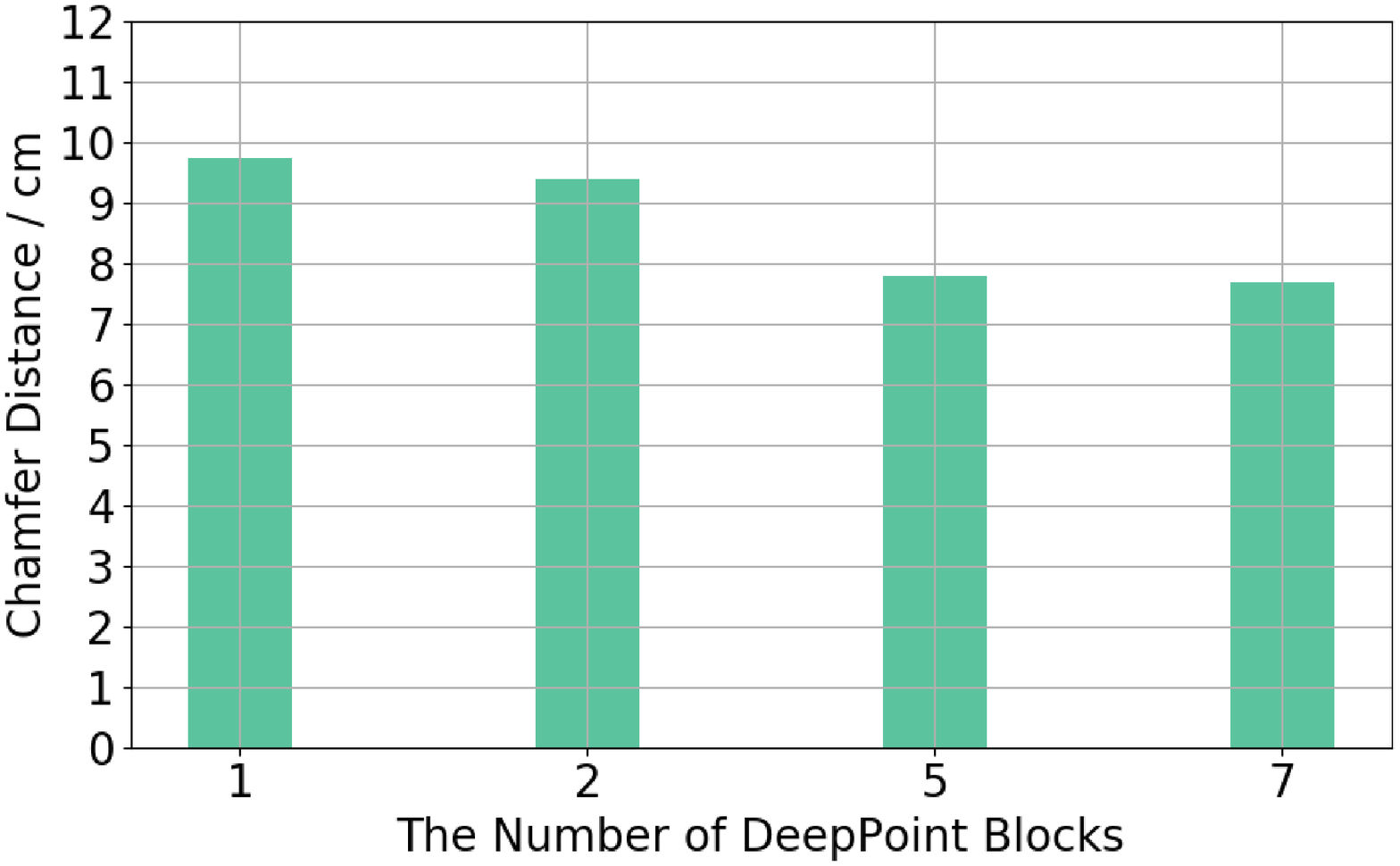}
				\setlength{\epsfxsize}{2.2in}
				\epsffile{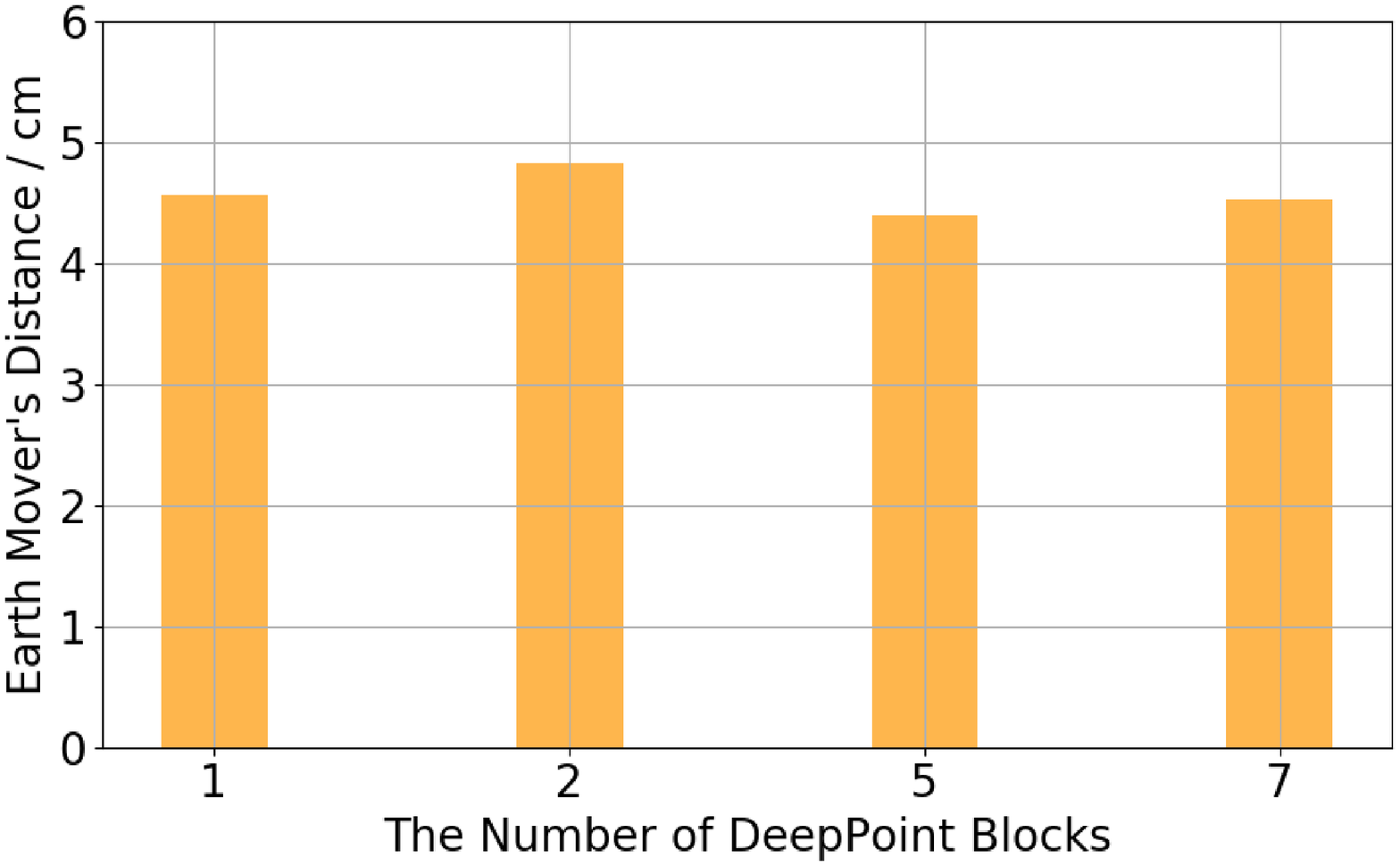}
				\setlength{\epsfxsize}{2.2in}
				\epsffile{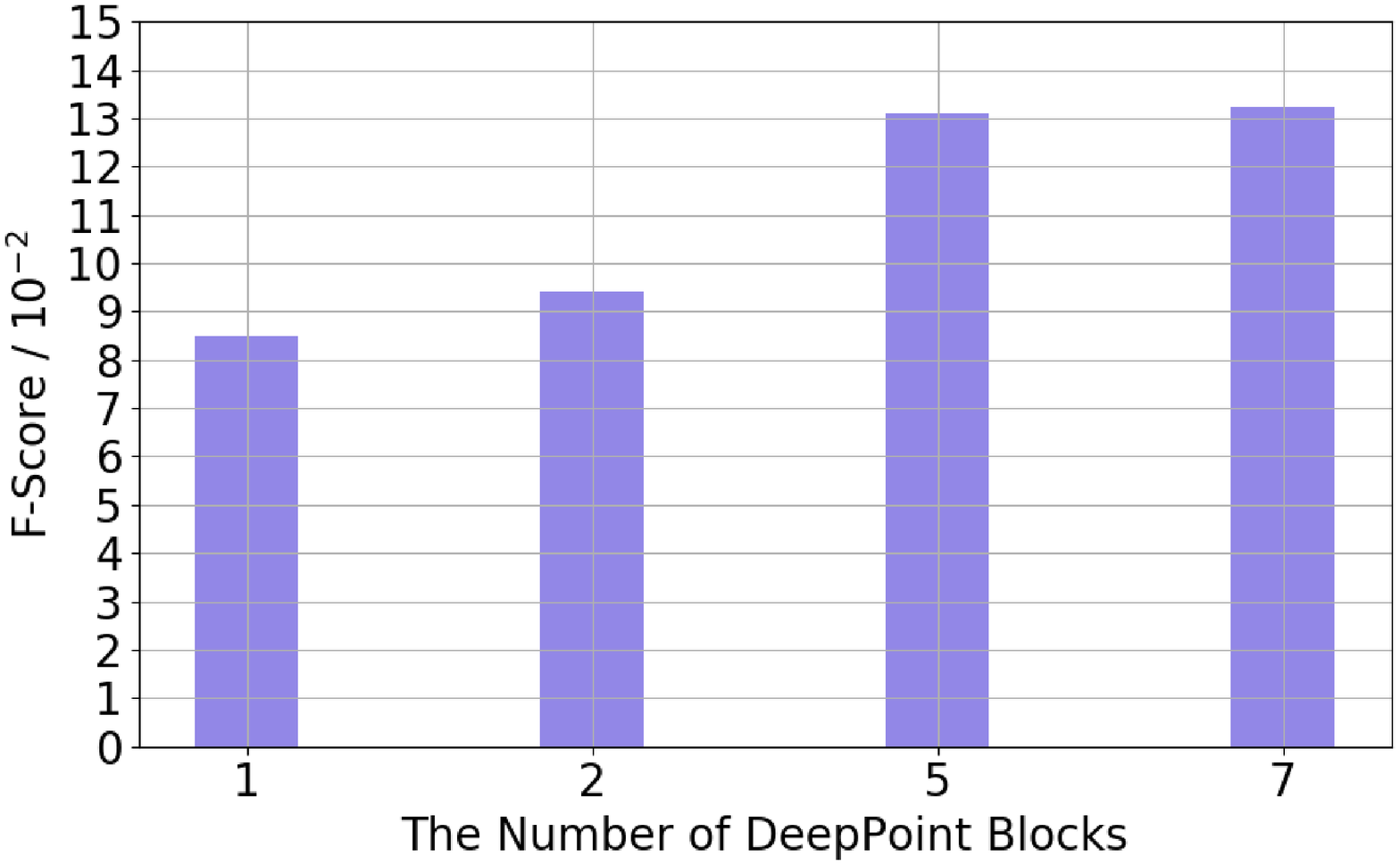}\\
				{}
			\end{center}
		\end{minipage}
	}
	\caption{Comparison of different numbers of DeepPoint blocks used in the design of the generator network.}
	\label{stage2_bar_pb}
\end{figure*}

The DeepPoint blocks in our generator network 
first expand each input point's dimension and then shrink them back to 3 to get each output point's coordinates.
The more such DeepPoint blocks means the generator network is ``deeper", 
which seems achieve better performance.
However, this is not always true. There exists an optimal number of layers. 
As shown in Fig. \ref{stage2_bar_pb},
the performance of our generator improves with the number of DeepPoint blocks increasing, 
i.e., CD decreases from 9.8 cm to 7.8 cm and EMD decreases from 4.6 cm to 4.4 cm as the number of DeepPoint Blocks increases from 1 to 5.
Correspondingly, F-score is even improved around $55\%$.
However, after the number of DeepPoint blocks reaches an upper bound, say 5 in our experiment,
further increasing the number of DeepPoint blocks to 7 can no longer largely improve the performance.
We can see that all these 3 evaluation metrics are very similar in these two cases.

\subsubsection{Skip Connections in Generator Network}
As shown in Fig. \ref{fig_stage_2}, we can see inside each DeepPoint block, 
we concatenate the raw input points' Cartesian coordinates 
with feature matrix obtained by passing through the shared MLP to further form the point features.
Our experiment results (e.g., Table \ref{tab_stage2_results_all}) clearly 
show that such skip connection design can improve the performance.
However, blindly increasing skip connections will not always help. 
As shown in another experiment in which we build 3 more skip connections by
concatenating the 1st and 7th point features, 2nd and 6th point features, 
and 3rd and 5th point features respectively.
However, based on the results shown in Table \ref{tab_stage2_results_all}, 
we can see that with these additional skip connections, generator performs worse compared with the case without using them.
In our future work, we will investigate an optimal placement of skip connections.
\subsubsection{Variants of Discriminator}
In our discriminator network design, we use mix pooling to extract the global feature vectors.
Mix pooling means that we concatenate the feature vectors from both max pooling and average pooling.
We compare the results of using max pooling, average pooling and mix pooling in the discriminator in Table \ref{tab_stage2_D}.
Note that the generators in these 3 experiments are the same. 
Table \ref{tab_stage2_D} shows that mix pooling performs best among these three pooling methods, 
and max pooling falls a little behind average pooling method.

\renewcommand{\arraystretch}{1.5} 
\begin{table}[tp]	
	\centering
	\fontsize{6.5}{8}\scriptsize
	\begin{tabular}{|c|c|c|c|c|c|c|}
		\hline
		\multirow{2}{*}{Method}&
		\multicolumn{2}{c|}{CD}&\multicolumn{2}{c|}{EMD}&\multicolumn{2}{c|}{ F-score}\cr\cline{2-7}
		&avg.&std.&avg.&std.&avg.&std.\cr
		\hline
		\hline
		Mix Pooling              &{\bf 9.75}   &4.00   &{\bf 4.56}    &3.96  &{\bf 8.47}    &3.44    \cr\hline
		Max Pooling             &10.52         &4.80   &4.79            &4.33  & 7.29           & 2.71        \cr\hline
		Average Pooling       &10.28         &4.11    &4.60           &4.11    &7.71            & 3.18          \cr\hline
	\end{tabular}
	\captionsetup{font={scriptsize}}
	\caption{Quantitative Results of Stage 2 using different pooling methods in the discriminator. Note that the units of CD and EMD in this table are cm, and the magnitude of F-score is $10^{-2}$.}
	\label{tab_stage2_D}
\end{table}

\section{CONCLUSIONS AND FUTURE WORK}\label{sec_conclusion}

We have proposed DeepPoint, a deep learning model 
that generates 3D objects in smooth and dense point clouds.
It utilizes a sequence of novel DeepPoint blocks to extract
essential features of the union of multiple rough and sparse 
input point clouds of an object when observed from
various viewpoints, even though the inputs contain many incorrect points. 
It relies on a deep structure design, an optimally chosen number 
of DeepPoint blocks, and skip connections to achieve 
good 3D reconstruction performance.
For future work, we will find the optimal placement of skip connections and introduce 
new techniques to improve the detailed geometry of generated point clouds. We will also conduct large scale experiments to improve our design.

%
%
\newpage

\bibliographystyle{IEEEtran}
\bibliography{IEEEabrv,references}

\end{document}